\documentclass[10pt,letterpaper]{article}
\usepackage[top=0.85in,left=2.75in,footskip=0.75in]{geometry}

\usepackage{amsmath,amssymb}

\usepackage{changepage}

\usepackage[utf8x]{inputenc}

\usepackage{textcomp,marvosym}

\usepackage{cite}

\usepackage{nameref,hyperref}

\usepackage{microtype}
\DisableLigatures[f]{encoding = *, family = * }

\usepackage[table]{xcolor}

\usepackage{array}

\usepackage{multirow}

\newcolumntype{+}{!{\vrule width 2pt}}

\newlength\savedwidth

\newcommand\thickhline{\noalign{\global\savedwidth\arrayrulewidth\global\arrayrulewidth 2pt}%
\hline
\noalign{\global\arrayrulewidth\savedwidth}}

\raggedright
\usepackage[aboveskip=1pt,labelfont=bf,labelsep=period,justification=raggedright,singlelinecheck=off]{caption}

\makeatletter
\renewcommand{\@biblabel}[1]{\quad#1.}
\makeatother

\usepackage{lastpage,fancyhdr,graphicx}
\fancyheadoffset[L]{2.25in}
\fancyfootoffset[L]{2.25in}
\newcommand{\firstuse}[1]{#1} 
\newcommand{\ie}{i.e.\ }

\newcommand{\etal}{et al.}
\newcommand{\ignore}[1]{}
\newcommand{\lstinline}[1]{\texttt{#1}} 
\newcommand{\heavy}[1]{\mathbf{#1}} 
\newcommand{\light}[1]{#1}

\newcommand{\E}{\mathrm{E}} 
\newcommand{\sd}{\mathrm{sd}} 
\DeclareMathOperator{\diag}{diag}
\DeclareMathOperator{\sign}{sign}
\newcommand{\abs}[1]{\vert#1\vert} 

\begin{document}
\vspace*{0.2in}

\begin{flushleft}
{\Large
\textbf\newline{ALAAMEE: Open-source software for fitting autologistic actor attribute models} 
}
\newline
\\
Alex Stivala\textsuperscript{1*},
Peng Wang\textsuperscript{2,1},
Alessandro Lomi\textsuperscript{3}
\\
\bigskip
\textbf{1} Institute of Computing, Universit\`a della  Svizzera italiana, Lugano, Switzerland
\\
\textbf{2} Centre for Transformative Innovation, Swinburne University of Technology, Melbourne, Australia
\\
\textbf{3} Universit\`a della  Svizzera italiana, Lugano, Switzerland
\\
\bigskip

%
%





* alexander.stivala@usi.ch

\end{flushleft}
\section*{Abstract}
The autologistic actor attribute model (ALAAM) is a model for social
influence, derived from the more widely known exponential-family random graph
model (ERGM). ALAAMs can be used to estimate parameters corresponding
to multiple forms of social contagion associated with network
structure and actor covariates. This work introduces ALAAMEE,
open-source Python software for estimation, simulation, and
goodness-of-fit testing for ALAAM models. ALAAMEE implements both the
stochastic approximation and equilibrium expectation (EE) algorithms for
ALAAM parameter estimation, including estimation from snowball sampled network
data. It implements data structures and statistics for undirected,
directed, and bipartite networks. We use a simulation study to assess
the accuracy of the EE algorithm for ALAAM
parameter estimation and statistical inference, and demonstrate the
use of ALAAMEE with empirical examples using both small (fewer than 100
nodes) and large (more than 10~000 nodes) networks.

\section*{Author summary}
If we observe a social network, along with some attributes of the
actors within it, including an opinion, behaviour, or belief (we will
call this the outcome attribute) that we might suppose to be socially
contagious, how might we test this supposition? The situation is
complicated by the fact that, if the outcome attribute is indeed
socially contagious, then its value for one actor will depend on its
value for that actor's network neighbours (friends, for instance).
Even further complexity arises if we suppose that the outcome is not
simply contagious like a disease, where it is likely to be transmitted
from a single network neighbour, but its adoption might instead
depend on particular patterns of adoption of the outcome attribute in
the local network structure surrounding an actor. The ``autologistic
actor attribute model'' (ALAAM) is a statistical model that, unlike
some better-known models, can handle such situations.  In this work we
describe open-source software called ALAAMEE that implements this
model, and demonstrate its use on both small and large networks.



\section*{Introduction}

Social contagion is a form of social influence supported by social contact.
Specifically, it
is the process whereby actors in a social network
adopt the attitudes, opinions, behaviours, or beliefs (which we refer
to generically as outcome attributes here) of their network neighbours. This
process may also be known as diffusion, one possible result of what is 
known in economics as peer effects~\cite{bramoulle20}. One source
of complexity in social networks and diffusion on them is that this
process results in autocorrelation in such outcome variables. That is,
the outcome variable is correlated across connected network nodes.
Such processes are therefore often inferred from observed data with
the network autocorrelation model
\cite{ord75,cliff81,doreian81,anselin90,friedkin90,leenders02}, in
which a parameter associated with network contagion is estimated.

Unlike a ``simple contagion'', such as, for example, disease, or
diffusion of information, in which a single tie to a network neighbour
with the outcome in question is likely sufficient for transmission, a
``complex contagion'' \cite{centola21} may require a certain threshold
of an actor's network neighbours to adopt a behaviour before that
actor is influenced to do so \cite{granovetter78,centola07,centola10}.
Such a process results in an even more complex relationship between
network structure, other attributes of the actors in the network, and
the outcome variable \cite{friedkin90,friedkin99,centola15}.

The autologistic actor attribute model (ALAAM) is a model for social
contagion which allows for such complexity. Unlike the more widely
known network autocorrelation model, in which social contagion is
associated with a single parameter \cite{doreian81,leenders02}, the ALAAM can be used to estimate
parameters corresponding to multiple mechanisms of social contagion
associated with more complex relationships between network structure,
the outcome variable, and other actor attributes. This is discussed
further in Stivala \etal~\cite{stivala20a}, and the dependency
assumptions and the types of terms these allow in the ALAAM are
discussed in Daraganova\cite{daraganova09} and Koskinen \&
Daraganova~\cite{koskinen22}.

Originally introduced by Robins \etal~\cite{robins01b}, the ALAAM is a
variant of the exponential-family random graph model (ERGM) for social
networks~\cite{robins01,daraganova09,daraganova13,lusher13,amati18,koskinen20,koskinen23}. As
such, the ALAAM is a cross-sectional model: given a single observation
of the social network and the outcome attribute (assumed to be binary in the
ALAAM) which is hypothesized to be socially contagious, the model is
used to estimate parameters relating to this contagion. In the ALAAM,
the network is assumed to be fixed (exogenous), and the binary outcome
variable is modeled (the variable is endogenous). This in contrast to
the ERGM, in which the tie variables are modeled (tie formation is an
endogenous process), based on fixed actor attributes. Note that the
outcome (binary) attribute of an actor in the ALAAM is allowed to
depend on its values for other actors connected to it in the social
network, hence the ``autologistic'' in the name.

ALAAMs can also be estimated for network data obtained via snowball
sampling~\cite{daraganova09,daraganova13,kashima13,stivala20a}. For
a recent introduction and review of ALAAM usage, see
Parker, Pallotti, \& Lomi~\cite{parker22},
and for a comprehensive survey of ALAAM applications,
Stivala~\cite{stivala23b}.

The most commonly used software for ALAAM modeling is the
IPNet Windows application~\cite{pnet}, or its successor,
the Windows application MPNet~\cite{mpnet14,mpnet22}, which allows
for ERGM and ALAAM modeling with undirected, directed, bipartite, and
multilevel networks. Note that the widely-used R package statnet~\cite{handcock08,hunter08,hummel12,ergm,ergm4} for ERGM
modeling, does not implement ALAAM.
Although, as noted in
Barnes \etal~\cite{barnes20}, an ALAAM can be considered as an ERGM for a two-mode
network, with the $N$ nodes of one mode representing the actors, and a
tie from an actor to the single node of the other mode representing
that actor having the outcome attribute, with the observed $N$ node
social network as a fixed covariate. And hence this could be
implemented as a statnet ergm model, but whether or not it is
practical is another question. This conception of an ALAAM could also
be a way of implementing a multivariate ALAAM, by having more than one
``outcome'' node.

The exponential-family random network model (ERNM), a generalization
of the ERGM and ALAAM, which models both social selection and social
contagion simultaneously~\cite{fellows12,fellows13,wang23} is
implemented using statnet. The only other publicly available code for
ALAAM modeling is the R code for estimating Bayesian ALAAMs described
by Koskinen \& Daraganova~\cite{koskinen22}.

These existing implementations are limited, both by the algorithms
implemented and details of their implementation, in the size of the
networks on which they can practically be used. Therefore, in order to
be able use ALAAMs with large (tens of thousands of nodes or more) networks,
such as, for example, those that can be collected from online social
network data, an alternative is required.

This paper describes one such alternative, \firstuse{ALAAMEE},
open-source Python code for ALAAM parameter estimation, simulation,
and goodness-of-fit testing. The ALAAMEE software implements the same
stochastic approximation algorithm for ALAAM parameter estimation as
IPNet and MPNet do~\cite{snijders02}, which is practical for networks
of the order of thousands, or in some cases tens of thousands, of
nodes in size or smaller. For larger networks, it also implements the
``equilibrium expectation'' (EE)
algorithm~\cite{byshkin16,byshkin18,borisenko19}, which has previously
been used for estimating ERGM parameters for very large
networks~\cite{byshkin18,stivala19_patent,stivala20b,stivala22}.
Because it implements both the stochastic approximation and EE
algorithms, the ALAAMEE software can be used to estimate ALAAM models
for both small and large (tens of thousands of nodes or more)
networks. It is not practical to do the latter with existing software,
and hence ALAAMEE extends the range of network data for which ALAAMs
are a practical modeling choice.

\section*{The autologistic actor attribute model (ALAAM)}

The ALAAM, modeling the probability of outcome attribute $Y$ (taking the form of
a binary vector $y$) given the network $X$ (a matrix of binary tie
variables) can be expressed as~\cite{daraganova13}:
\begin{equation}
  \label{eqn:alaam}
  \mathrm{Pr}_{\theta}(Y = y \vert X =x) = \frac{1}{\kappa(\theta)}\exp\left(\sum_I \theta_I z_I(y,x,w)\right)
\end{equation}
where $\theta_I$ is the parameter corresponding to the
network-attribute statistic $z_I$, in which the ``configuration'' $I$
is defined by a combination of dependent (outcome) attribute variables $y$,
network variables $x$, and actor covariates $w$, and
$\kappa(\theta)$ is the normalizing quantity which ensures a proper
probability distribution.

Just as for ERGM, parameter estimation for ALAAM is a computationally
intractable problem, and so the maximum likelihood estimate (MLE) of
the ALAAM parameters is found using Markov chain Monte Carlo (MCMC)
methods (see Hunter \etal~\cite{hunter12} for an overview in the ERGM context).

The sign and significance of the ALAAM parameters $\theta_I$ support
inferences about the statistical relationship between the
corresponding configurations $z_I(y,x,w)$ and the outcome attribute
binary vector $y$, each conditional on all the other effects included
in the model.

\begin{figure}
  \centering
  \includegraphics[width=\textwidth]{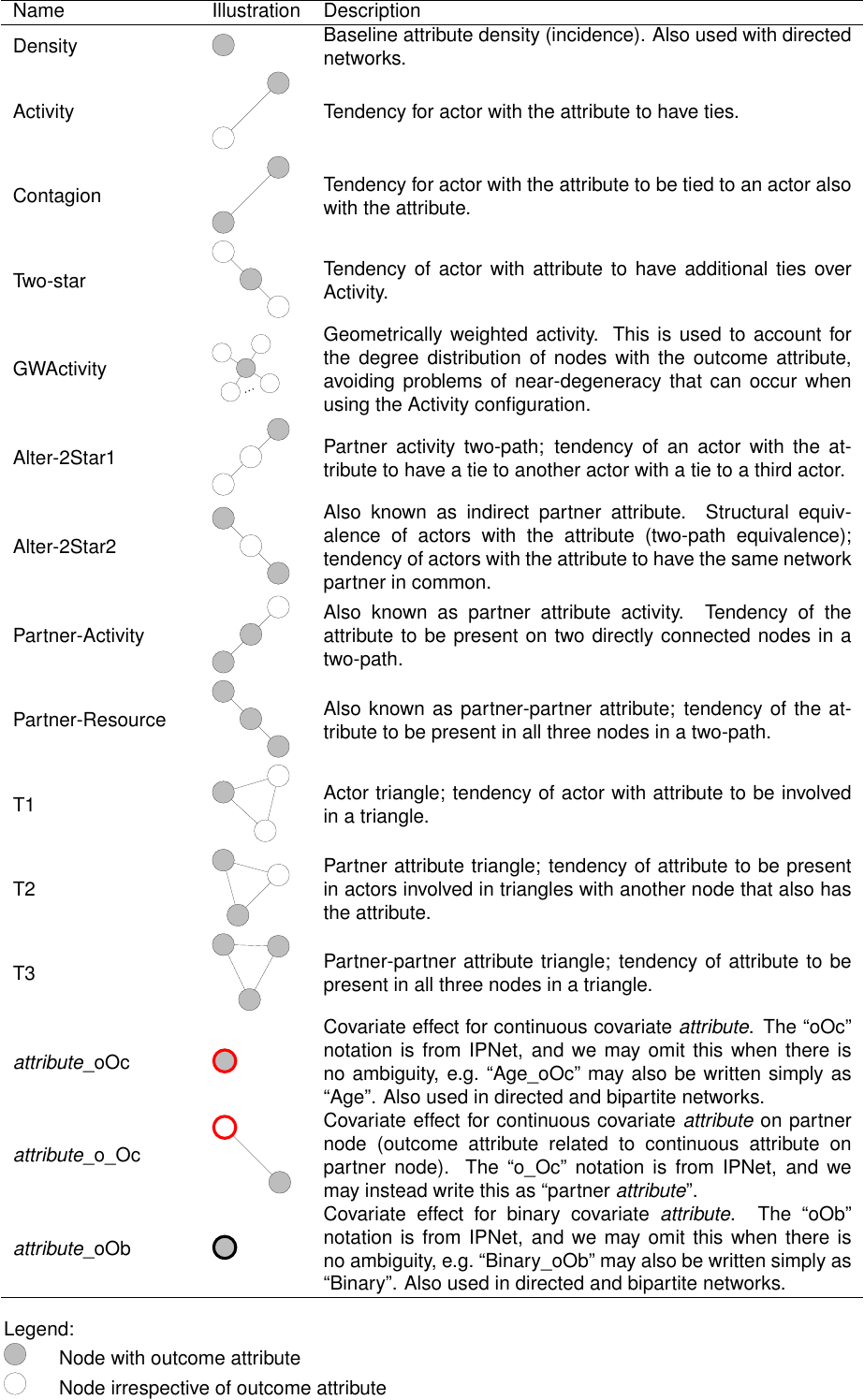}
\caption{{\bf Configurations used in ALAAMs for undirected networks.}}
\label{fig:changestats_undirected}
\end{figure}

For example, consider the frequently used contagion effect, using an
undirected network for simplicity. The statistic for contagion is the
number of pairs of directly connected nodes, $i$ and $j$ ($i \neq j$)
where $x_{ij} = 1$, in which both nodes have the outcome attribute
$y_i = y_j = 1$ (see Fig~\ref{fig:changestats_undirected}). If the
contagion parameter (that is, the parameter corresponding to the contagion statistic just defined) is found to be positive and significant, this
means the contagion configuration (a directly connected pair of nodes
both with the outcome attribute) occurs more frequently than expected
by chance, given all the other effects included in the model. So there
is a statistically significant correlation between a pair of nodes
being directly connected, and both having the outcome attribute
(conditional on the other effects in the model).

The distinction between the ERGM and the ALAAM arises from the
assumptions and hypotheses behind the model. In the ERGM, modeling tie
formation, we might hypothesize that having the same attribute value
affects (increases, for homophily) the probability of a tie between
two nodes, and test for this. In the ALAAM, modeling the outcome
attribute, we might hypothesize that having a network tie affects
(increases, for contagion) the probability that both nodes will have
the outcome attribute.  More generally, the difference between the ERGM
and the ALAAM reflects the distinction introduced by Borgatti \&
Halgin~\cite{borgatti11} between theories of networks and network
theories. The former concern the possible antecedents of network
ties. The latter concern the potential outcomes of network ties.
Distinguishing homophily and contagion in an observational network
study is in general a difficult problem~\cite{shalizi11}, and ALAAM
parameters indicate only (auto\nobreakdash-)correlation, not
causation.

Once parameters for an ALAAM model of an observed network and outcome
binary attribute have been estimated, these can be used to simulate,
again using MCMC, a set of outcome binary vectors from the model.
These simulated outcome attributes can be used for goodness-of-fit testing and
model evaluation.
For further explanation and examples, see Parker, Pallotti, \& Lomi~\cite{parker22}.

\section*{Sampling from the ALAAM distribution}
\label{sec:sampling}

Simulating outcome attribute vectors from a model specified by a parameter
vector $\theta$ requires drawing from the ALAAM distribution defined
by Eq~(\ref{eqn:alaam}). Drawing from this distribution is also
necessary for the estimation algorithms (described in the following
section). Due to the intractable normalizing constant
$\kappa(\theta)$ in Eq~(\ref{eqn:alaam}), the Metropolis--Hastings
algorithm is used for this purpose. This MCMC algorithm proposes
transitions from the current outcome binary vector $y$ to a proposed
vector $y'$ with acceptance probability
\begin{align}
  p_\alpha &= \min\left\{1, \frac{q(y' \rightarrow y) \Pr_\theta(Y = y' \vert X =x)}{q(y \rightarrow y') \Pr_\theta(Y = y \vert X =x)} \right\} \\
         &= \min\left\{1, \frac{q(y' \rightarrow y)}{q(y \rightarrow y')} \exp\left(\sum_I \theta_I \delta_I(y,x,w)\right)\right\}   \label{eqn:mh_acceptance}
\end{align}
where $q$ is the proposal distribution and
$\delta_I(y,x,w) = z_I(y',x,w) - z_I(y,x,w)$
is the \firstuse{change statistic}~\cite{hunter06,snijders06,hunter12}
corresponding to the statistic $z_I$ for each configuration $I$ in the model.
Note that this means that only these change statistics $\delta$ need to be
computed, rather than the statistics $z$ themselves.

If the proposal distribution is symmetric, that is 
$q(y' \rightarrow y) = q(y \rightarrow y')$, then Eq~(\ref{eqn:mh_acceptance})
can be simplified to
\begin{equation}
  p_\alpha = \min\left\{1,  \exp\left(\sum_I \theta_I \delta_I(y,x,w)\right)\right\}   \label{eqn:metropolis}
\end{equation}
and this is an instance of the Metropolis, rather than the more
general Metropolis--Hastings, algorithm.

The simplest choice of proposal, which we refer to as the basic ALAAM
sampler, is to choose a node uniformly at random and toggle its
outcome attribute value, that is, change it to 1 if it was 0,
and to 0 if it was 1. In the former case the change statistic is
$\delta_I(y,x,w)$  and in the latter case it is
$-\delta_I(y,x,w)$. This proposal distribution is symmetric and
so the simplified Metropolis acceptance probability, Eq~(\ref{eqn:metropolis}),
can be used.

In the case of ERGMs, the analogous basic sampler (toggling a dyad)
often results in very low acceptance rates, due to the sparsity of the
network, and more efficient samplers (with asymmetric proposal
distributions) such as the TNT (``tie / no tie'') sampler
\cite{morris08} or IFD (improved fixed density) sampler
\cite{byshkin16} are frequently used. In the case of ALAAM, however,
the frequency (density or incidence) of the outcome variable is
typically orders of magnitude higher the graph density of a sparse
graph such as those to which ERGMs are usually applied, and so more
advanced samplers are not necessary.  We implemented an ALAAM analogue
of the TNT sampler, the ZOO (``zero or one'') sampler, but found that
it had little or no practical application (it did not speed up the process
of sampling from the ALAAM distribution). The basic ALAAM sampler is
used for all experiments and examples in this work.

\section*{Estimation algorithms}

\subsection*{Stochastic approximation (SA)}
\label{sec:sa}

Stochastic approximation of ERGM parameters using the Robbins--Monro
algorithm~\cite{robbins51} is described in detail in
Snijders~\cite{snijders02}, and recapitulated more briefly in Koskinen
\& Snijders~\cite{koskinen13b} and very briefly in Hunter \etal~\cite{hunter12}.
Here
we briefly describe this algorithm and its application to the estimation
of ALAAM parameters. As this is a relatively straightforward
application of the same algorithm in a slightly different context, the
preceding references can referred to for further detail.

The aim of the algorithm is to find the MLE of the parameter vector $\theta$, which solves
\begin{equation}
\nabla_\theta \log \mathrm{Pr}_\theta(Y = y) = z(y) - \E_\theta(z(Y)) = 0 \label{eqn:moment}
\end{equation}
where $\nabla_\theta \log \Pr_\theta(Y = y)$ is the gradient of the
log-likelihood $\log \Pr_\theta(Y = y)$ \cite{hunter12} (but note here
$Y$ and $y$ are random binary outcome attribute vectors and instances of binary
outcome attribute vectors, respectively, not matrices of tie variables as used in
ERGM).  That is, we aim to find the parameter vector $\theta$ such
that $\E_\theta(z(Y))$, the expected value of the statistics with
respect to the probability distribution (\ref{eqn:alaam}), is equal
to the observed statistics $z(y) =  z(y_{\mathrm{obs}}) = z_{\mathrm{obs}}$.

The SA algorithm to do this consists of three phases. Phase 1
estimates the covariance matrix based on a small number of samples
$M$ sampled from the ALAAM distribution at the initial parameter
value $\theta^{(0)}$ as
\begin{equation}
  D = \frac{1}{M} U^\mathsf{T} U \label{eqn:estD}
\end{equation}
where $U$ is a matrix such that each row
$U_{i,\cdot} = Z_{i,\cdot} -\overline{z}$
where $Z$ is a matrix where each row is the
vector of statistics for a sample and $\overline{z}$ is the vector
of mean statistics over the $M$ samples. We use $M = M_1 = 7 + 3p$ where $p$ is the number of parameters in the model.

Phase 2 is the main part of the algorithm, where the parameter
estimates $\theta$ are updated using the Robbins--Monro algorithm according
to:
\begin{equation}
  \theta^{(t+1)} = \theta^{(t)} - a_i  D^{-1} (z^{(t)} - z_{\mathrm{obs}})
\end{equation}
where $z^{(t)} = z \left(y^{(t)} \right)$ is the vector of statistics of a sample $y^{(t)}$
drawn from the ALAAM
distribution with parameter vector $\theta^{(t)}$ and $a_i$ is a sequence
of positive step sizes converging to zero, satisfying the
Robbins--Monro convergence criteria: at each subphase this value is
halved. We use five subphases and $a_0 = 0.01$.

Phase 3 estimates the covariance matrix of the estimator to estimate
the standard error, checks the approximate validity of
Eq~(\ref{eqn:moment}) and tests for convergence. This is done by
estimating the covariance matrix $D$ just as in phase 1 using
Eq~(\ref{eqn:estD}), but with a larger number of samples $M = M_3 = 1000$
drawn from the ALAAM distribution at the final estimated value of $\theta$.
If this matrix is singular then the model may be degenerate.
Otherwise the standard error is estimated as
\begin{equation}
  \mathrm{se}(\theta) = \sqrt{\diag\left(D^{-1}\right)} \label{eqn:sa_stderr}
\end{equation}
and the t-ratio as
\begin{equation}
  t_i = \frac{\overline{z}_i - z_{\mathrm{obs},i}}{\sd(z_i)}
\end{equation}
for each parameter $i$ in the model. If the absolute value of the t-ratio $\abs{t_i} < 0.1$ for each parameter $i$
then the estimation is considered to be converged: the difference
between the expected value of the statistic under the model with the estimated parameters $\theta$ and its observed value
is small enough that we consider Eq~(\ref{eqn:moment}) to be satisfied.
If this convergence criterion is not satisfied, then the entire algorithm
may be run again, starting at the current estimated value of the
parameters $\theta$.

In this algorithm, whenever samples are drawn from the ALAAM
distribution (using the Metropolis--Hastings algorithm described in
section ``\nameref{sec:sampling}'') an interval of $10N$
iterations, where $N$ is the number of nodes, is used between each
sample to reduce autocorrelation between samples. In phase 3, a
burn-in period of $M_3 N$ iterations is used to try to ensure that it
has reached equilibrium before samples are taken.

\subsection*{Equilibrium expectation (EE)}
\label{sec:ee}

The stochastic approximation algorithm is a robust and reliable
algorithm that does not require particularly good initial estimates
\cite{koskinen13b}. However, it has the shortcoming that it is not
scalable to very large networks since it involves running the MCMC
algorithm for sampling from the ALAAM distribution to convergence each
time the parameter estimates are updated.

The EE algorithm \cite{byshkin16,byshkin18,borisenko19} overcomes this
scalability problem by not requiring potentially very long MCMC
simulations between parameter updates \cite{stivala20b}. Instead, it
attempts to find the MLE, that is, solve Eq~(\ref{eqn:moment}), by
taking only a small number ($M_s = 1000$) of steps of the
Metropolis--Hastings algorithm (see section ``\nameref{sec:sampling}'')
between updates of the estimated parameter vector $\theta$ according
to the divergence of the simulated statistics from the observed
statistics. After many iterations (default value
$M_{\textrm{EE}} = 50000$) of this process, if on average this divergence is zero, then we
have estimated $\theta$ values for the MLE. Note that in the EE
algorithm, the Markov chain of simulated outcome attribute vectors starts at the
observed value (not, for example, a zero vector or random vector),
and the Markov chain continues across iterations of the EE algorithm
(that is, the simulated outcome attribute vector is not reset after the parameters are updated).

The parameter update step in the EE algorithm is
\begin{equation}
  \theta^{(t+1)} = \theta^{(t)}  -\sign(z^{(t)} - z_{\mathrm{obs}}) \cdot r \cdot \max\left\{\abs{\theta^{(t)}}, c \right\} \label{eqn:ee_update}
\end{equation}
where
$z^{(t)} = z \left( y^{(t)} \right)$ 
is the vector of statistics at the current state of the
simulated outcome attribute vector $y^{(t)}$ (initially $z^{(0)} = z_{\mathrm{obs}}$),
$r > 0$
is the learning rate ($r = 0.01$ by default) and $c = 0.01$ is a
constant that ensures the algorithm does not get ``stuck'' at zero
values of the parameters.
The parameter update is actually performed as
\begin{equation}
  \theta^{(t+1)} = \theta^{(t)}  -\sign(d_z^{(t)}) \cdot r \cdot \max\left\{\abs{\theta^{(t)}}, c \right\} \label{eqn:ee_update_impl}
\end{equation}
where $d_z^{(t+1)} = d_z^{(t)} + \delta_M$ (initially, $d_z^{(0)} = 0$),
and $\delta_M$ are the change statistics for accepted moves summed over the $M_s$ iterations
of the Metropolis--Hastings algorithm in this EE iteration.
Since $d_z$ accumulates change statistics for accepted Metropolis--Hastings moves, starting at zero for the
observed outcome attribute vector, then $d_z^{(t)} = z^{(t)} - z_\mathrm{obs}$ and
so Eq~(\ref{eqn:ee_update_impl}) is equivalent
to Eq~(\ref{eqn:ee_update}), and $z_\mathrm{obs}$ need not be computed.

Note that the version of the EE algorithm used here is the simplified
version of Borisenko \etal~\cite{borisenko19} rather than the original
version \cite{byshkin18,stivala20b}.

This algorithm results in chains of $\theta^{(t)}$ and $d_z^{(t)}$ values,
where $0 \leq t < M_\textrm{EE}$. To compute the point estimate and
standard error for $\theta$ we first thin the chains by
discarding the first 1000 iterations (burn-in) and
using an interval of 100 iterations between each value, resulting
in $N_m = (M_\textrm{EE} - 1000) / 100$ values for each.

Similar to the situation discussed in Hunter \& Handcock
\cite{hunter06} for ERGM estimation, there are two sources of error that need to be
considered: the MCMC error in our point estimate of $\theta$, and the
error inherent in using the MLE.
The point estimate for $\theta$ and its asymptotic covariance
matrix $T$ are estimated by the multivariate batch means
method~\cite{dai17,flegal10,jones06,vats18,vats19} from the thinned
$\theta^{(t)}$ chain, as is the asymptotic covariance matrix for the
simulated statistics, $V$, from the thinned $d_z^{(t)}$ chain.
The estimated standard errors for the $\theta$ estimates are
then estimated by
\begin{equation}
  \mathrm{se}(\theta) = \sqrt{\diag(W)}
\end{equation}
where
\begin{equation}
  W = \frac{1}{N_m} T + \left(\frac{1}{N_m} V\right)^{-1}.
\end{equation}

In doing the computations described above, several convergence checks
can be applied. If the covariance matrix $V$ is computationally
singular, then the model is possibly degenerate, and this estimation
is not valid. The estimation can also be discarded as non-converged
if, heuristically, it appears not to converge due to $\theta$ values
that are infinite or NaN (``not a number'') or simply ``too large''
(heuristically, $\abs{\theta}> 10^{10}$), or if the average of the
$d_z$ values after the burn-in period is not approximately zero
(for this purpose we use the heuristic that an estimation is
considered non-converged if $\abs{\overline{d_z} / \sd(d_z)} > 0.3$).

Unlike the stochastic approximation algorithm (but not unlike the MCMC MLE
algorithm used for ERGM parameter estimation in statnet
\cite{hunter06,handcock08,hunter08,hummel12,ergm,ergm4,koskinen13b,krivitsky17}),
it is important to have a reasonable initial estimate of the
parameters for the EE algorithm, and for that purpose we use
``Algorithm S'' from the EE algorithm publications
\cite{byshkin18,stivala20b}, which, as discussed in those papers and
also the paper describing the simplified EE algorithm
\cite{borisenko19} used here, is equivalent to contrastive divergence
(CD) \cite{hinton02}, which has been applied to ERGM parameter
estimation \cite{asuncion10,fellows14,krivitsky17}. Essentially this
algorithm amounts to running a small number (we use $100$ here) of the
EE parameter update steps (\ref{eqn:ee_update}), but without actually
performing any accepted moves in the Metropolis--Hastings algorithm
(that is, leaving the outcome attribute vector at the observed value).

When using the original version of this algorithm to estimate ERGM
parameters \cite{byshkin18,stivala20b}, it was found that it tended to
have low statistical power because of large estimated standard errors.
For this reason, a meta-analysis technique, similar to that used for
combining ERGM parameter estimates from multiple snowball samples of
large networks \cite{stivala16} is used. Specifically, the EE
algorithm is run multiple times (which can be done in parallel, taking
advantage of modern multicore or parallel computing clusters), and the
estimates combined by the inverse-variance weighted
average~\cite[Ch.~4]{hartung08}. From the vector of $N_C$ (the
number of converged estimates) estimated parameter values $\theta$ and
a vector of their associated estimated standard errors $s$, the
point estimate is
\begin{equation}
  \hat{\theta} = \frac{\sum_{j=1}^{j=N_C} \theta_j / s_j^2}{\sum_{j=1}^{j=N_C} 1 / s_j^2}
\end{equation}
and the estimated standard error is
\begin{equation}
  \mathrm{se}(\hat{\theta}) = \frac{1}{\sqrt{\sum_{j=1}^{j=N_C} 1/ s_j^2}}
\end{equation}
for each of the parameters in the model.
In this work we use this method, running multiple EE algorithm
runs in parallel for each ALAAM estimation.

In order to test that the parameter estimates are properly converged
and the model is not degenerate or near-degenerate, ALAAM vectors can be simulated
from the model with estimated parameter vector $\hat{\theta}$ and the
statistics of these vectors compared to those of the observed outcome
attribute vector, to verify that Eq~{(\ref{eqn:moment})} is (approximately)
satisfied. Here we do this by plotting degeneracy check plots, showing
trace plots and histograms of the simulated vector statistics, along
with the observed values, from which it can be verified that the
simulation has sufficient burn-in and large enough interval to avoid
excessive autocorrelation between samples, and that the observed
values are within the 95\% confidence interval of sample
distribution. Note that, unlike the EE algorithm itself, these
simulations require sufficient burn-in and iterations to ensure the
simulation has reached the stationary distribution and the samples are
not too autocorrelated.  For very large networks, therefore, this can mean
that this degeneracy check simulation could possibly take longer than
the estimation with the EE algorithm.

If the estimation does not converge, it may be useful to reduce the
value of the learning rate $r$, which affects how large
the parameter update steps are. As discussed in Giacomarra
\etal~\cite{giacomarra23}, it needs to be sufficiently small for the
EE algorithm to converge to the MLE, but a smaller learning rate
$r$ may come at the expense of requiring a larger number of iterations
$M_\textrm{EE}$. For example, in estimating ALAAM parameters for the
very large (approximately 1.6 million node) network in
Stivala~\cite{stivala23b}, the learning rate was set to $r =
0.001$. For all estimations in this work, the default values of the
learning rate ($r = 0.01$) and iterations ($M_\textrm{EE} = 50000)$
were used.

\section*{Materials and methods}
\label{sec:methods}

\subsection*{Implementation}

ALAAMEE is implemented in Python 3 \cite{python}, and uses the NumPy~\cite{harris20} package for array data types and linear algebra. The
Python code does not require any other packages, simply using a
``dictionary of dictionaries'' data structure to implement graphs.
Just as described in Bianchi \etal~\cite{bianchi22}, this allows simple and
efficient graph construction and implementation of the graph
operations required, such as testing for the existence of an edge or
arc, and iterating over the neighbours of a node.

ALAAM parameters are usually estimated by stochastic approximation
with the Robbins--Monro algorithm~\cite{robbins51,snijders02}. This is
the method used in MPNet, and is also implemented in ALAAMEE,
as described in the ``\nameref{sec:sa}'' section.
For the stochastic approximation algorithm, all stages of the
estimation, including point estimation, estimation of standard errors
from the Fisher information matrix~\cite{snijders02}, and
simulation-based goodness-of-fit testing, are implemented in Python.

For larger networks, on the order of tens of thousands of nodes or
more, this method may no longer be practical. For such networks, the
EE algorithm can be used, and ALAAMEE also implements this algorithm,
as described in the ``\nameref{sec:ee}'' section.  The EE algorithm
works differently~\cite{stivala20b}. In this algorithm, a number of
estimation processes (each of which is a separate Python task) are run
independently. Because these runs are independent, they can be run in
parallel to minimize the elapsed time taken.

From the results of these multiple estimation runs, a point estimate
and estimated standard error are computed using an R \cite{R-manual} script that uses
the mcmcse R package~\cite{mcmcse} to estimate the Fisher information
matrix and the asymptotic covariance matrix for the MCMC
standard error with the multivariate batch means method~\cite{dai17,flegal10,jones06,vats18,vats19}. The overall
estimate and its estimated standard error are then computed from these
multiple independent runs as the inverse variance weighted average~\cite[Ch.~4]{hartung08}.

Scripts are provided to run the parallel estimations using either GNU
Parallel~\cite{tange18} or, for Linux compute clusters, SLURM~\cite{yoo03} job arrays. Scripts for processing network data and
converting it between different formats, taking snowball samples from
networks, making plots, and computing the Wilson score interval~\cite{wilson27} for the binomial proportion score interval (used to
find confidence intervals for Type I and II error rates in the
simulation study described in the ``\nameref{sec:results}'' section) are
written in R and use the igraph~\cite{csardi06}, ggplot2~\cite{ggplot2}, and PropCIs~\cite{propcis} R packages.

ALAAMEE also implements estimation (and simulation) conditional on
snowball sampled network structure, as described in
Pattison \etal~\cite{pattison13} and Stivala \etal~\cite{stivala16} in the context of ERGM, and in
Daraganova~\cite{daraganova09}, Daraganova \& Pattison~\cite{daraganova13b}, Kashima \etal~\cite{kashima13}, and Stivala \etal~\cite{stivala20a} for ALAAM.

\subsection*{Change statistics}

It is a property of the ALAAM
that the odds of a node having the outcome attribute equal to one,
conditional on the values of the outcome attribute for the other nodes, is a
function of the change in the vector of statistics associated with
switching the outcome attribute value of that node from 0 to 1. It
follows that, in implementing the MCMC process for ALAAM simulation
and estimation, only these change statistics need be implemented
(see section ``\nameref{sec:sampling}'' for details).
That is,
rather than writing functions to count each of the configurations in
the data, we need only write functions that compute the change
statistic value resulting from changing the outcome attribute value of
a given node from 0 to 1. The value of the statistics in observed data
can be computed by summing the change statistics for each element of
the outcome attribute vector that is equal to 1.

For example, consider the simplest statistic, attribute
\firstuse{Density}, sometimes instead called \firstuse{Incidence}~\cite{parker22} to avoid confusion with graph density. This statistic
is simply the number of nodes with the outcome attribute equal to 1. Its
corresponding change statistic is simply the constant 1, since for any
node, if its outcome attribute value is switched from 0 to 1, then
that increases the Density statistic by 1.

\begin{figure}
  \centering
  \includegraphics[width=0.8\textwidth]{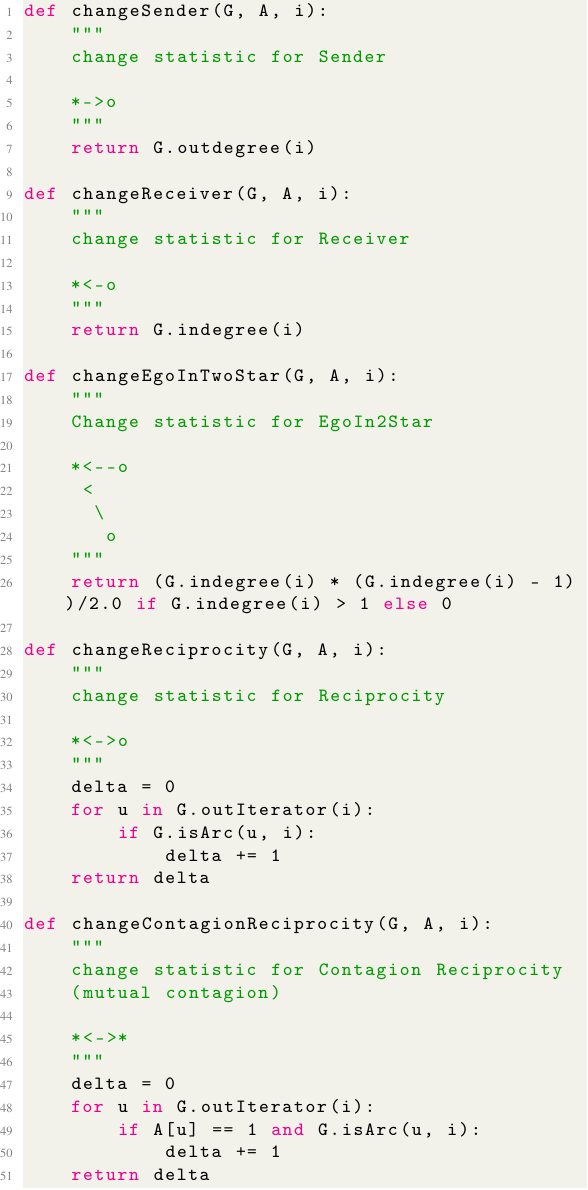}
  \caption{{\bf Example change statistic implementations.} Some change statistics for directed networks.}
\label{fig:change_stat_example_code}
\end{figure}

Every change statistic in ALAAMEE is implemented as a function of the
form \lstinline{change\textit{Statname}(G, A, i)} where
\lstinline{G} is a \lstinline{Graph} (or \lstinline{Digraph} or
\lstinline{BipartiteGraph}, as appropriate to the change statistic)
object, \lstinline{A} is a binary outcome attribute vector, and \lstinline{i} is
a node identifier. The function returns the change statistic value for
switching the node outcome attribute value \lstinline{A[i]} from 0 to
1. It is a precondition of the change statistic functions that
\lstinline{A[i] == 0}. Some examples are shown in
Fig~\ref{fig:change_stat_example_code}. Python ``docstrings'' are
used to document the change statistic functions, including ASCII
diagrams illustrating the corresponding configurations. This allows the
documentation to be viewed at the interactive Python prompt with the
built-in \lstinline{help()} function.

\begin{figure}
  \centering
  \includegraphics[width=\textwidth]{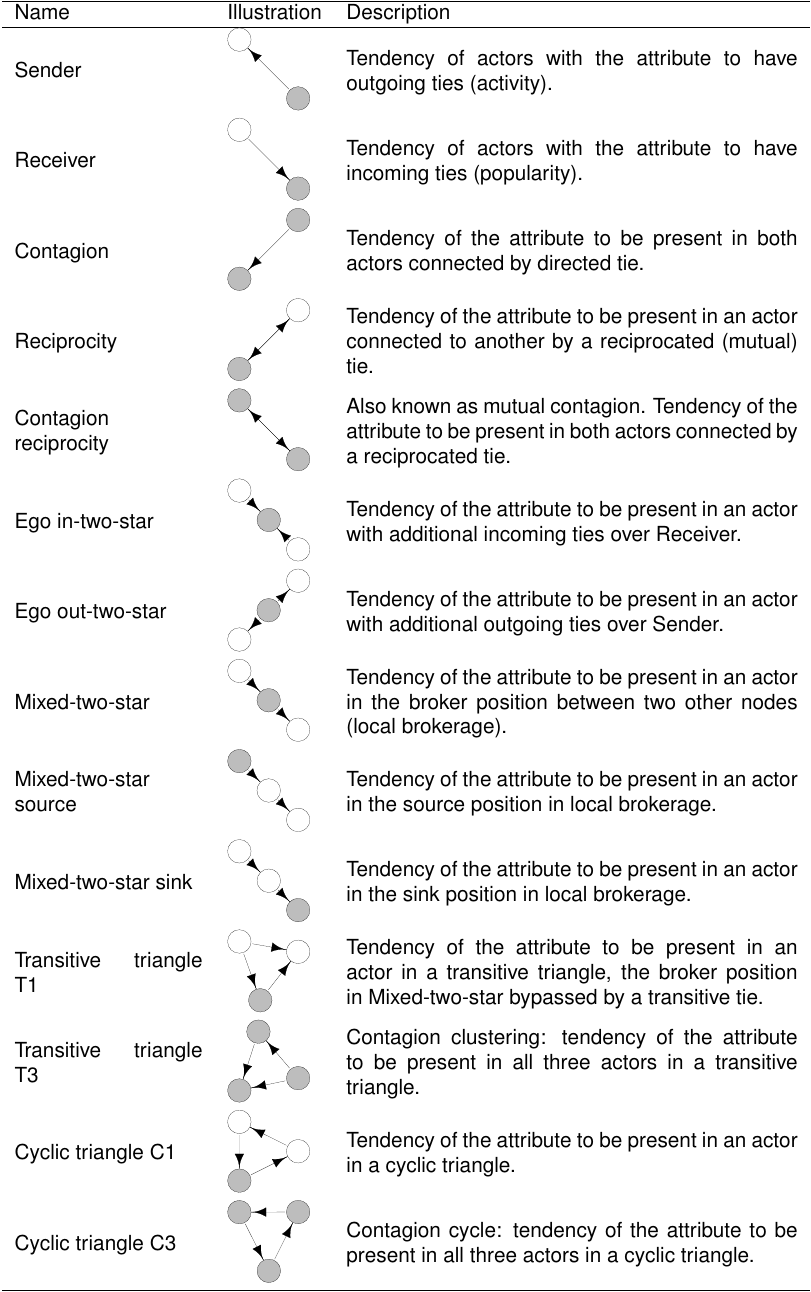}
  \caption{{\bf Configurations used in ALAAMs for directed networks.}}
\label{fig:changestats_directed}
\end{figure}

The first two functions shown in
Fig~\ref{fig:change_stat_example_code} implement, respectively, the
\firstuse{Sender} and \firstuse{Receiver} change statistics (see
Fig~\ref{fig:changestats_directed}). These are quite
straightforward. The Sender statistic counts the number of outgoing
ties from actors that have the outcome attribute equal to 1. Therefore, the
change statistic when a node $i$ has its outcome variable changed from
0 to 1 is the out-degree of $i$, and similarly for Receiver and
in-degree.

The third function, \lstinline{changeEgoInTwoStar}, is slightly more
complicated. This statistic (see Fig~\ref{fig:changestats_directed})
counts the number of pairs of incoming arcs to nodes that have the
outcome attribute equal to 1. Hence the change statistic when a node $i$ has its
outcome variable changed from 0 to 1 is the number of pairs of
incoming arcs to node $i$, that is
$\binom{d^{\mathrm{(in)}}_i}{2} = d^{\mathrm{(in)}}_i (d^{\mathrm{(in)}}_i - 1) / 2$,
where $d^{\mathrm{(in)}}_i$ is the in-degree of node $i$ (and $d^{\mathrm{(in)}}_i > 1$).

The last function shown in Fig~\ref{fig:change_stat_example_code} implements
the change statistic for the \firstuse{Contagion reciprocity}
statistic, also known as \firstuse{mutual contagion} (see
Fig~\ref{fig:changestats_directed}). This statistic counts the
number of pairs of nodes with a reciprocated (mutual) tie between
them, in which both nodes have the outcome attribute equal to 1. Hence the change
statistic, computing the change in the statistic when a node $i$ has
its outcome variable changed from 0 to 1, counts the number of
out-neighbours of $i$, \ie nodes $u$ such that there is an arc $i
\rightarrow u$, such that $u$ has the outcome variable equal to 1 and there is
also an arc $u \rightarrow i$. \lstinline{changeReciprocity} is similar, but does not require that both nodes have the outcome attribute equal to 1, only that node $i$ does.
(Note that the code for \lstinline{changeContagionReciprocity} in Fig~\ref{fig:change_stat_example_code} can be
written less verbosely and more elegantly, and arguably in more
idiomatic Python style, in a single line using a list comprehension:
\lstinline{return sum([(G.isArc(u, i)and A[u] == 1) for u in G.outIterator(i)])},
however this turns out to be slower; similar considerations apply to \lstinline{changeReciprocity}.)
The ALAAMEE source code
repository (\url{https://github.com/stivalaa/ALAAMEE})
includes unit tests for verifying the correctness of change statistic
implementations against known correct values,
verifying that alternative implementations give the same results, and comparing
their execution speeds.

Some change statistics make use of nodal attribute values. These
change statistic functions take also as their first parameter the
name of the attribute to use, used as the key in the relevant
attribute dictionary in the \lstinline{Graph} object. So that these
functions have the same signature as the structural statistics, the
higher-order function \lstinline{functools.partial()} is used to
create a function with the \lstinline{(G, A, i)} signature. For
example, \lstinline{partial(changeoOc, "age")} returns a function with
the required \lstinline{(G, A, i)} signature, implementing the change
statistic for outcome attribute related to nodal continuous attribute
\lstinline{"age"}, given the original generic change statistic
function \lstinline{changeoOc(attrname, G, A, i)}. This usage is
illustrated in the empirical example in the ``\nameref{sec:results}'' section.
The same technique is used for statistics that use an auxiliary
(``setting'') network (that is, statistics that depend on edges in a second network, as used for example in Divi\'ak \etal\cite{diviak20}, where the network is criminal collaboration, and the setting network is pre-existing ties such as kinship or friendship), a distance matrix, the decay value for
geometrically weighted statistics~\cite{stivala23b}, and for the node
type (mode) in two-mode graphs using the \lstinline{BipartiteGraph}
object.

For some change statistics, computation can be made far more efficient
and scalable by using a sparse matrix counting two-paths between each
pair of nodes in the network. A similar technique, implementing the
sparse matrix as a hash table, has been used for ERGM change
statistics~\cite{stivala20b}. This technique is far simpler and more
advantageous for ALAAM estimation and simulation in ALAAMEE, for two
reasons. First, in ALAAM the network is exogenous, and
hence, once constructed, the two-path lookup sparse matrix need not be
modified. Second, the sparse matrix data structure is very easily and
efficiently implemented in Python as a ``dictionary of dictionaries''
data structure, just as the graph data structures are, since
dictionaries are a built-in data type in Python.

Currently, change statistics for undirected and directed one-mode
networks, and undirected two-mode (bipartite) networks are
implemented. These statistics include all of those used in the ALAAM
models published in Kashima \etal~\cite{kashima13}, Letina~\cite{letina16}, Letina \etal~\cite{letina16b}, and Divi{\'a}k \etal~\cite{diviak20} for
undirected networks, and Gallagher~\cite{gallagher19} and Parker \etal~\cite{parker22} for directed
networks, for example. ALAAMEE has been used to estimate ALAAM
parameters for a director interlock (bipartite) network~\cite{stivala23}. Change statistics for the new geometrically
weighted ALAAM statistics described in Stivala~\cite{stivala23b} are
also implemented in ALAAMEE.

We hope that the use of Python will facilitate the implementation
of further user-defined change statistics, since adding ALAAM change
statistics to the Python code, for example as shown in
Fig~\ref{fig:change_stat_example_code}, would appear to be
considerably easier than the procedure for adding new
ERGM change statistics in the statnet ergm package, which requires
writing both R and C code~\cite{hunter13}.

\subsection*{Data sources}
\label{sec:data}

\subsubsection*{Simulation study}
\label{sec:simulation}

To evaluate the performance of the EE algorithm
implemented in ALAAMEE for ALAAM parameter estimation and statistical
inference, we apply it to estimating parameters for ALAAM outcome attribute
vectors with known parameters. These are obtained by generating the
outcome attribute vectors by ALAAM simulation from a fixed network (and
covariates) with a given set of ALAAM parameters. This is done over a
set of 100 simulated ALAAMs, allowing us to measure the point estimate
bias and RMSE (root mean square error), as well as the type I (false
positive) and type II (false negative) error rates in inference.

This technique was used to evaluate ERGM estimation from snowball
sampled network data in Stivala \etal~\cite{stivala16} and for ERGM estimation by
the EE algorithm in Stivala \etal~\cite{stivala20b}. It was used to
evaluate ALAAM estimation from sampled network data in
Stivala \etal~\cite{stivala20a}.

Here we use the simulated ALAAM outcome attribute values on the ``Project 90''
network, a sexual contact network of high-risk heterosexuals in
Colorado Springs~\cite{potterat04,woodhouse94,klovdahl94,rothenberg95}. These are
exactly the simulations described in Stivala \etal~\cite{stivala20a}. The network
is the giant component of the Project 90 network, consisting of 4430
nodes, with mean degree $8.31$. Binary and continuous attributes are
generated for the nodes: the binary attribute is assigned the nonzero
value for 50\% of the nodes, chosen at random, and the continuous
attribute value $v_i$ at each node $i$ is $v_i
\stackrel{iid}{\thicksim} N(0,1)$. From this network and nodal
attributes, ALAAM outcome attribute vectors are simulated with parameters
(Density, Activity, Contagion, Binary, Continuous) =
$(-15.0, 0.55, 1.00, 1.20, 1.15)$. As described in Stivala \etal~\cite{stivala20a},
these parameters were chosen so that approximately 15\% of the nodes have
the outcome attribute value equal to 1.

\subsubsection*{Small network}

To demonstrate the implementation of the stochastic
approximation algorithm for estimating ALAAM parameters, we will use
the excerpt of 50 girls from the ``Teenage friends and lifestyle
study'' data~\cite{michell97,pearson00,pearson03,pearson06,steglich06,west96},
which is used as an illustrative example for the SIENA software~\cite{rsiena} for stochastic actor-oriented models (SAOMs)~\cite{snijders17}. The data consists of an excerpt of 50 girls from
panel data recording friendship networks and substance use over a
three year period from 1995, when the pupils were aged 13, to 1997, in
the West of Scotland. The data includes information on smoking
(tobacco), alcohol, and cannabis consumption, as well as sporting
activities.

This data was also used as a tutorial example for the Bayesian
ALAAM~\cite{koskinen22} implementation in R \cite{koskinen_github_alaam}.
As noted in the description for this
data \cite{s50data},
this is not a properly delineated network, and is used only for
illustrative purposes.

Unlike SAOMs, we cannot use ALAAMs to model the co-evolution of network
and actor covariates based on longitudinal data: we can only model a
single binary outcome attribute given a fixed network and other (fixed)
covariates, as well as the outcome attribute itself on other nodes. As in the
Bayesian ALAAM R tutorial, we will use smoking at the second wave as
the outcome variable, and use the network and other covariates from
the first wave. Hence the network assumed to diffuse social contagion
is observed before the outcome variable, as discussed in
Parker \etal~\cite{parker22}.

The categorical variable for smoking is converted to binary by treating
any amount of smoking other than completely non-smoking as the
nonzero binary outcome.

\subsubsection*{Large networks}

For networks with tens of thousands of nodes or more, estimating ALAAM
parameters with the stochastic approximation algorithm may no longer
be practical. For such networks, we can instead use the
EE algorithm. We will use three such networks as examples for
ALAAMEE parameter estimation: undirected online friendship networks
for the ``Deezer'' music streaming service in Croatia, Hungary, and
Romania~\cite{rozemberczki19}. These networks are publicly available
from the Stanford large network dataset collection~\cite{snapnets}.
Descriptive statistics of these networks are shown in
Table~\ref{tab:deezer_network_stats}.

\begin{table}[!ht]
\begin{adjustwidth}{-2.25in}{0in} 
\centering
\caption{
{\bf Network descriptive statistics for the Deezer networks.}}
\begin{tabular}{|l|r|r|r|r|r|r|r|}
\hline
Network & Nodes  &   Mean    & Max.     &    Density & Clustering   & Likes     & Likes             \\
&        &   degree  & degree   &            & coefficient  & jazz \%  & alternative \%   \\ \thickhline
Deezer Croatia & 54573 & 18.26 & 420 & 0.00033 & 0.11463 & 5 & 38\\ \hline
Deezer Hungary & 47538 & 9.38 & 112 & 0.00020 & 0.09292 & 5 & 37\\  \hline
Deezer Romania & 41773 & 6.02 & 112 & 0.00014 & 0.07527 & 6 & 36\\  \hline
\end{tabular}
\begin{flushleft} Each network is a single connected
    component. Network statistics were computed using the igraph R
    package. ``Clustering coefficient'' is the global
    clustering coefficient (transitivity).
\end{flushleft}
\label{tab:deezer_network_stats}
\end{adjustwidth}
\end{table}

Each node in these three networks represents a user, and an undirected
edge represents friendship in the Deezer online social network. Each
node is annotated with a list of genres liked by the user~\cite{rozemberczki19}. Based on these genre annotations, we created
two different binary outcome variables: one for liking jazz, and one
for liking ``alternative'' music. The binary outcome variable for
liking jazz is true if the user likes any one or more of the genres in
the data that describe jazz music, namely ``Jazz'', ``Instrumental jazz'',
``Jazz Hip Hop'', or ``Vocal Jazz''. The binary outcome variable for
liking alternative music is true if the user likes the genre in the
data labelled ``Alternative''. Consistently across all three networks,
approximately 5\% of users like jazz, and between 35\% and 40\% of
users like alternative music (Table~\ref{tab:deezer_network_stats}).

\section*{Results and Discussion}
\label{sec:results}

\subsection*{Simulation study for estimation using the equilibrium expectation algorithm}

\begin{figure}
  \centering
  \includegraphics[width=\textwidth]{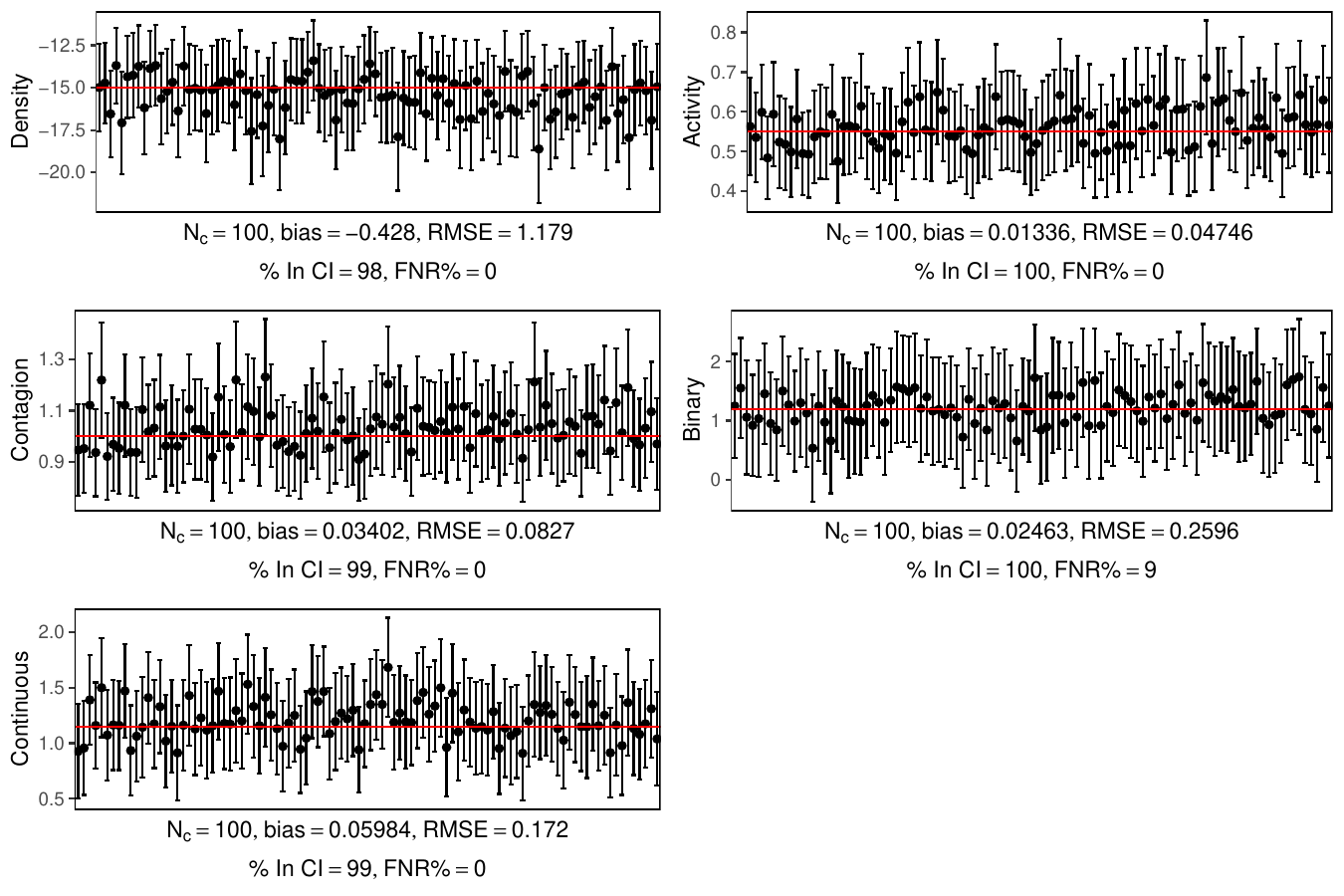}
  \caption{{\bf Parameter estimates and estimated standard errors
    from the EE algorithm.}
    The algorithm was used to estimate the known ALAAM parameters for the Project 90 network with simulated attributes. The error
    bars show the nominal $95\%$ confidence interval. The horizontal
    line shows the true value of the parameter, and each plot is
    annotated with the mean bias, root mean square error (RMSE), the
    percentage of samples for which the true value is inside the
    confidence interval (coverage rate), and the Type II error rate (False Negative
    Rate, FNR). $\mathrm{N_C}$ is the number of samples (of the total
    100) for which a converged estimate was found.}
\label{fig:meanse}
\end{figure}

Fig~\ref{fig:meanse} and Table~\ref{tab:fnr} show the results for
using the EE algorithm implemented in ALAAMEE to estimate parameters
from the simulated ALAAMs. For all parameters other than Binary (the
covariate effect for the simulated binary attribute described under
``\nameref{sec:simulation}'' in the ``\nameref{sec:data}'' section), the type II
error rate was less than 5\%. For the Binary parameter, however, the
type II error rate was estimated to be 9\%, with a 95\% confidence
interval $[5\%, 16\%]$. When using the stochastic approximation
algorithm on the same data, used as the baseline for comparing against
results from sampled data in Stivala \etal~\cite{stivala20a}, the type
II error rate was less than 5\% for all parameters, including Binary
(\nameref{sifig:ipnet_meanse}). Comparing the results for the EE
algorithm (Fig~\ref{fig:meanse}) and the stochastic approximation
algorithm (\nameref{sifig:ipnet_meanse}) for the Binary parameter, it
seems that the problem is likely not the point estimate, but rather
that the estimated standard errors (and hence confidence interval)
from the EE algorithm were too large, giving a high type II (false
negative) error rate on the Binary parameter. Although this problem
only occurs on the Binary parameter, the coverage rate (the percentage
of samples for which the true value is inside the confidence interval)
was higher than the nominal 95\% on all the parameters: in fact was
between 98\% and 100\% for the EE algorithm, while for the stochastic approximation
algorithm, it was still higher than 95\%, but less than 99\%
(\nameref{sifig:ipnet_meanse}).

\begin{table}[!ht]
\begin{adjustwidth}{-2.25in}{0in} 
\centering
\caption{
{\bf Type II error rate from estimation of simulated outcomes using the EE algorithm.}}
\begin{tabular}{|l|r|r|r|r|r|r|r|r|r|}
\hline
Effect &  Bias &  RMSE &  \multicolumn{3}{c|}{False negative rate (\%)}  & in C.I.    & Total     & Mean       & Total\\
\cline{4-6}
&       &       &   Estim. & \multicolumn{2}{c|}{95\% C.I.}       &  (\%)      & samples  & runs       & runs per\\
\cline{5-6}
       &       &       &   & lower & upper                              &            & converged & converged  & sample\\
\thickhline
Density  & -0.4280 & 1.1790 &  0 &  0 &  4 & 98 & 100 & 200.00 & 200\\ \hline
Activity  & 0.0134 & 0.0475 &  0 &  0 &  4 & 100 & 100 & 200.00 & 200\\ \hline
Contagion  & 0.0340 & 0.0827 &  0 &  0 &  4 & 99 & 100 & 200.00 & 200\\ \hline
Binary  & 0.0246 & 0.2596 &  9 &  5 & 16 & 100 & 100 & 200.00 & 200\\ \hline
Continuous  & 0.0598 & 0.1720 &  0 &  0 &  4 & 99 & 100 & 200.00 & 200\\ \hline
\end{tabular}
\begin{flushleft}  The ``estim.'', ``lower'', and ``upper''
    columns show the point estimate and lower and upper 95\%
    confidence interval (C.I.), respectively, of the Type II error
    rate (false negative rate). This C.I. is computed as the Wilson
    score interval. The ``in C.I. (\%)''
    column is the coverage rate for the nominal 95\% confidence
    interval. Results are over 100 simulated ALAAM outcome attribute vectors
    (samples), each of which is estimated with 200 parallel estimation
    runs.
\end{flushleft}
\label{tab:fnr}
\end{adjustwidth}
\end{table}

In order to measure the type I error rate on each parameter, an ALAAM
simulation without the corresponding effect (the parameter value is
zero) was required. Hence for each of the ALAAM parameters considered
(other than Density, which if zero results in almost all nodes having
the outcome attribute equal to 1), another set of 100 ALAAM outcome attribute vectors were
simulated, with the parameter set to zero and the other parameters
unchanged (apart from Activity, which if zero results in very few
nodes having the outcome attribute equal to 1, and so Density was changed to
$-7.0$ for this case only).

The results for estimating the type I error rate for ALAAM estimation
by the EE algorithm are shown in
Table~\ref{tab:fpr}. In all cases the type I error rate was less than
the nominal 5\%. Just as for the type II error rate results, however,
the coverage rate was higher than the nominal 95\%, indicating that the
standard error estimates might be too large.

\begin{table}[!ht]
\begin{adjustwidth}{-2.25in}{0in} 
\centering
\caption{
{\bf Type I error rate from estimation of simulated outcomes using the EE algorithm.}}
\begin{tabular}{|l|r|r|r|r|r|r|r|r|r|}
\hline
Effect &  Bias &  RMSE &  \multicolumn{3}{c|}{False positive rate (\%)}  & in C.I.   & Total     & Mean       & Total\\
\cline{4-6}
&       &       &   estim. & \multicolumn{2}{c|}{95\% C.I.}       &  (\%)      & samples  & runs       & runs per\\
\cline{5-6}
&       &       &   & lower & upper                              &            & converged & converged  & sample\\ \thickhline
Activity  & 0.0048 & 0.0149 &  0 &  0 &  4 & 100 & 100 & 200.00 & 200\\ \hline
Contagion  & 0.0011 & 0.0386 &  0 &  0 &  4 & 100 & 100 & 200.00 & 200\\ \hline
Binary  & 0.0191 & 0.2922 &  1 &  0 &  5 & 99 & 100 & 200.00 & 200\\ \hline
Continuous  & 0.0475 & 0.1638 &  0 &  0 &  4 & 100 & 100 & 200.00 & 200\\ \hline
\end{tabular}
\begin{flushleft}  The ``estim.'', ``lower'', and ``upper''
    columns show the point estimate and lower and upper 95\%
    confidence interval (C.I.), respectively, of the Type I error
    rate (false positive rate). This C.I. is computed as the Wilson
    score interval. The ``in C.I. (\%)''
    column is the coverage rate for the nominal 95\% confidence
    interval. Results are over 100 simulated ALAAM outcome attribute vectors
    (samples), each of which is estimated with 200 parallel estimation
    runs.
\end{flushleft}
\label{tab:fpr}
\end{adjustwidth}
\end{table}

\begin{figure}
  \centering
  \includegraphics[width=\textwidth]{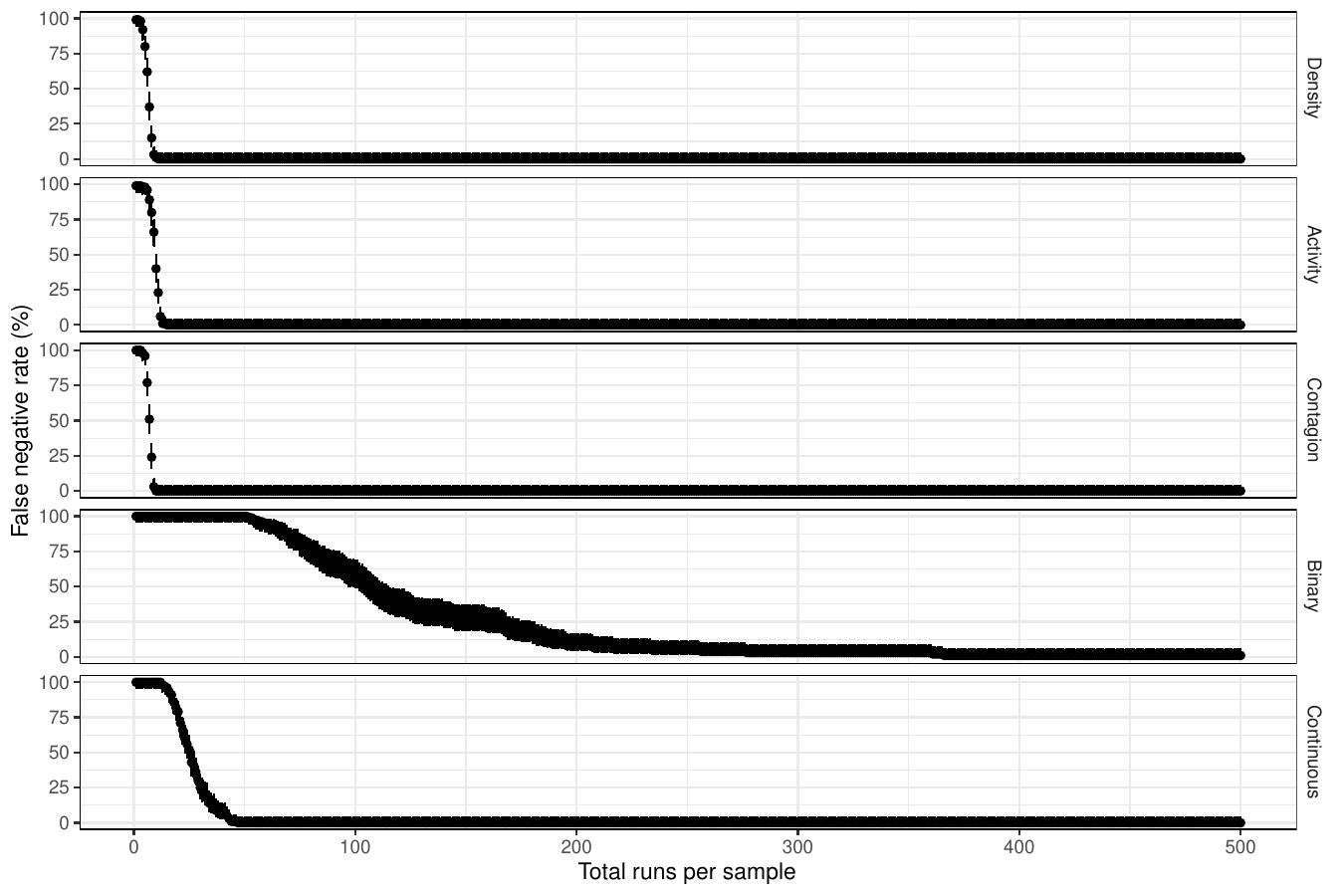}
  \caption{{\bf Type II error rate as the number of runs used for each sample is varied.}}
\label{fig:fnr_var_total_runs}
\end{figure}

In the results discussed so far, the point estimates and standard
errors were estimated, as described in the ``\nameref{sec:methods}'' section,
from 200 parallel estimation runs. Fig~\ref{fig:fnr_var_total_runs}
shows the effect on the type II error rate when the number of runs was
varied from 1 up to 500 (the data point for 200 runs therefore
corresponding to the results shown in Table~\ref{tab:fnr} and
Fig~\ref{fig:meanse}). This shows that for all parameters other than
Binary, 50 runs (indeed, for the Density, Activity, and Contagion
parameters, fewer than 20 runs) were more than sufficient to
obtain a type II error rate of less than 5\%. For the Binary
parameter, however, the type II error rate declined far more slowly,
and was 9\% with 200 runs, as we have seen.

Each of these estimation runs, with the default EE algorithm
parameters, took on average 49 minutes (sd = 12.7, min.~= 44, max.~=
126, median = 46, N = 8000) on an Intel Xeon E5-2650 v3 2.30 GHz
processor on a Linux cluster, and on average 9.46 minutes (sd = 0.49,
min.~= 3.59, max.~=12.85, median = 9.56, N = 40000) on AMD EPYC
7543 2.8 GHz processors on a newer and larger Linux cluster.
Less than 2 GB of memory was required
(for all parallel runs occupying a single node) for the Python code,
and less than 8 GB in total for the R scripts to do the final
statistical analyses and diagnostic plots.

\nameref{sifig:fpr_var_total_runs} shows the corresponding results for
the type I error rate. It shows that it is ``safe'' to vary the number of runs,
in the sense that the type I error rate remained less
than 5\% at all values tested.

The results of these simulation experiments indicate that it is desirable
for the purposes of decreasing the type II error rate, without
increasing the type I error rate to an unacceptable level, to use as many runs of the ALAAMEE
estimation process as practical (at least up to a maximum of 500). Because these runs are independent,
they can be run in parallel to minimize elapsed time, taking advantage
of as many processor cores as may be available. The results here
indicate that 200 runs is a reasonable number, however for some
parameters this might still result in an undesirably high type II
error rate.

\subsection*{Empirical examples}

\subsubsection*{Small network}

\begin{figure}
  \centering
  \includegraphics[width=\textwidth]{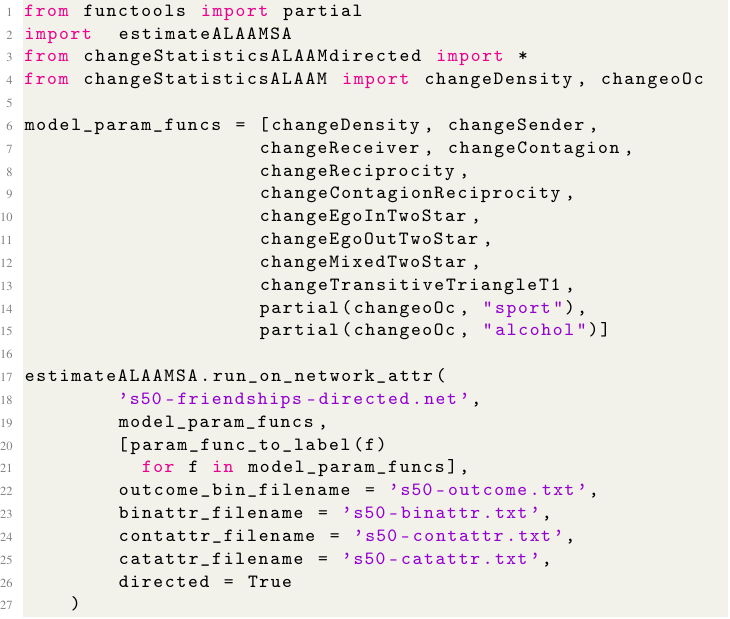}
  \caption{{\bf Python code for estimating parameters of an ALAAM
      model.} This code uses ALAAMEE with the stochastic
    approximation algorithm to estimate ALAAM parameters and their
    standard errors for the teenage friends and lifestyle data
    excerpt, with smoking as the outcome variable. After a converged
    model is found, this will also do a goodness-of-fit test.}
\label{fig:s50_code}
\end{figure}

\begin{figure}
  \centering
 \includegraphics[width=0.8\textwidth]{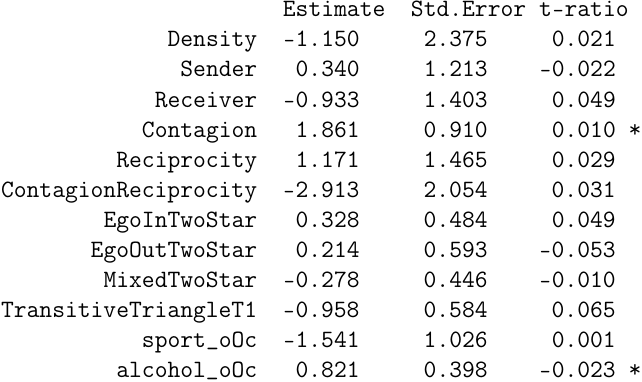}
\caption{{\bf Example ALAAMEE output.} This is the output from using
  ALAAMEE to estimate ALAAM parameters by stochastic approximation for
  the teenage friends and lifestyle data excerpt, with smoking as the
  outcome variable. Asterisks indicate statistical significance at $p
  < 0.05$.}
\label{fig:s50_results}
\end{figure}

Python code to use ALAAMEE to specify an ALAAM model for the excerpt
of 50 girls from the ``Teenage friends and lifestyle study'' and
estimate its parameters using the stochastic approximation
algorithm~\cite{snijders02} is shown in Fig~\ref{fig:s50_code} (this
code is located in the ALAAMEE repository as
\url{https://github.com/stivalaa/ALAAMEE/blob/master/examples/simple/directed/glasgow_s50/runALAAMSAs50.py}). It
also automatically does a goodness-of-fit test for the converged model
(if found). This code took less than two minutes (and less than 300 MB
of memory) to run on a Windows 10 personal computer with an Intel Core
i5-10400 2.90 GHz processor, giving the model shown in
Fig~\ref{fig:s50_results}. The results are consistent (only the
contagion and alcohol effects are significant, and positive) with
those from estimating the same model using MPNet~\cite{mpnet14}. For
this small network, MPNet was faster, taking approximately one minute
(and less than 100 MB of memory), on the same PC.

The model shown in Fig~\ref{fig:s50_results} indicates a significant
and positive contagion effect: smokers tend to be directly connected
to other smokers. The only other significant effect is for alcohol
consumption; this is positive, indicating that drinkers are more
likely to smoke.

\begin{figure}
  \centering
  \includegraphics[width=0.5\textwidth]{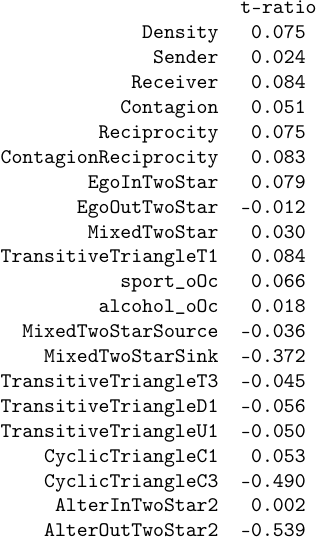}
  \caption{{\bf Example ALAAMEE goodness-of-fit output.} This is the
    goodness-of-fit output from ALAAMEE for the model it estimated by
    stochastic approximation for the teenage friends and lifestyle
    data excerpt, with smoking as the outcome variable.}
\label{fig:s50_gof}
\end{figure}

The goodness-of-fit statistics for this model are shown in
Fig~\ref{fig:s50_gof}. The effects in the model are at the top, followed
by additional effects which are not included in the model. For effects
in the model, the convergence statistic (t-ratio) should be less than
0.1 in magnitude (just as they must be for the model to be considered
converged by the estimation algorithm). Fig~\ref{fig:s50_gof}
shows that all the effects included in the model met this criterion.

For effects not included in the model, a rule of thumb is that the
t-ratio should be less than $2.0$ in magnitude for an acceptable fit~\cite{parker22}, although some authors have used a stricter threshold
of $1.0$~\cite{kashima13,daraganova13b}, or even $0.3$~\cite{diviak20}.
Parker \etal~\cite{parker22} use $1.645$, signifying statistical
significance at the $5\%$ level (one-tailed). All of the effects not
included in the model, but included in the goodness-of-fit test shown
in Fig~\ref{fig:s50_gof}, were less than $1.0$ in magnitude, indicating
this model had an acceptable fit for all the statistics included
in the goodness-of-fit test.

\subsubsection*{Large networks}

Table~\ref{tab:deezer_jazz_ee_estimations} shows the results of using
the EE algorithm implemented in ALAAMEE to estimate parameters for
ALAAM models of the three Deezer online social networks, with liking
jazz as the outcome variable. The model parameters were selected to
test for social contagion of liking jazz on this social network (the
Contagion parameter), including specifically in closed triads (the
three Triangle parameters), while also accounting for the degree
distribution of users who like jazz (the GWActivity parameter \cite{stivala23b}), and
the number of genres liked by a user and their friends.  Degeneracy
check plots for these models are shown in 
\nameref{sifig:deezer_hr_degen_check}, \nameref{sifig:deezer_hu_degen_check},
and \nameref{sifig:deezer_ro_degen_check}.

\begin{table}[!ht]
\begin{adjustwidth}{-2.25in}{0in} 
\centering
\caption{
{\bf Models estimated using ALAAMEE for the Deezer networks, with liking jazz as the outcome variable.}}
\begin{tabular}{|l|r|r|r|}
  \hline
  Effect  & Croatia & Hungary & Romania\\ \thickhline
Density  & $\heavy{\underset{(-4.866, -4.589)}{-4.727}}$ & $\heavy{\underset{(-5.209, -4.901)}{-5.055}}$ & $\heavy{\underset{(-4.870, -4.513)}{-4.691}}$\\ \hline
GWActivity [$\alpha = 2.0$]  & $\light{\underset{(-7.280, 1.390)}{-2.945}}$ & $\heavy{\underset{(3.269, 7.200)}{5.234}}$ & $\light{\underset{(-0.835, 4.047)}{1.606}}$\\ \hline
Contagion  & $\light{\underset{(-0.130, 0.115)}{-0.008}}$ & $\heavy{\underset{(0.153, 0.377)}{0.265}}$ & $\heavy{\underset{(0.403, 0.823)}{0.613}}$\\ \hline
TriangleT1  & --- & --- & $\light{\underset{(-0.076, 0.008)}{-0.034}}$\\ \hline
TriangleT2  & --- & --- & $\light{\underset{(-0.177, 0.185)}{0.004}}$\\ \hline
TriangleT3  & --- & --- & $\light{\underset{(-0.749, 0.710)}{-0.020}}$\\ \hline
num. genres  & $\heavy{\underset{(0.186, 0.201)}{0.194}}$ & $\heavy{\underset{(0.184, 0.201)}{0.193}}$ & $\heavy{\underset{(0.184, 0.203)}{0.194}}$\\ \hline
partner num. genres  & $\light{\underset{(-0.001, 0.001)}{-0.000}}$ & --- & $\heavy{\underset{(-0.011, -0.001)}{-0.006}}$\\ \hline
\end{tabular}
\begin{flushleft} Parameter estimates are
    shown with 95\% confidence interval. Estimates that are
    statistically significant at the nominal $p < 0.05$ level are
    shown in bold. Results are from 200 converged runs (of a total 200).
\end{flushleft}
\label{tab:deezer_jazz_ee_estimations}
\end{adjustwidth}
\end{table}

This model was estimated first on the Romania network (the smallest of
the three). Initially the Activity parameter, rather than GWActivity,
was used, however the degeneracy check showed that this model was not
properly converged, and hence the geometrically weighted activity
parameter GWActivity (with decay parameter $\alpha = 2.0$) was used
instead of Activity to overcome this problem, as described in
Stivala~\cite{stivala23b}. This model, however, did not converge for the
Croatia and Hungary data, but removing the triangle parameters solved
this problem, resulting in the models shown in
Table~\ref{tab:deezer_jazz_ee_estimations} (for Hungary, the partner
number of genres parameter also had to be removed to find a converged
model).

Density (or incidence) is negative and statistically significant in
all models. This effect, analogous to the intercept in logistic
regression, is simply the baseline presence of the outcome attribute
(liking jazz). Its negative value merely reflects the fact that jazz
is a minority interest in the data for all three countries.

The only parameter, apart from Density, that is significant across all
three networks is the number of genres liked by a user: the more
genres a user likes, the more likely they are to like jazz. This could
simply be a consequence of the fact that, even if we assume liking
jazz is completely random, the more genres someone likes, the more
likely it is that jazz is one of them (simply by chance), and this
parameter controls for this effect. It also has a more substantive
interpretation: we might suppose that a preference for jazz is a
highly specific taste that excludes also liking other genres, and so
people who like jazz are unlikely to also like other genres. The
negative and statistically significant estimate of this parameter
refutes that hypothesis.

The Contagion parameter is positive and
significant for both Hungary and Romania, indicating that (given the
assumptions of the model) a liking for jazz is socially contagious in
these countries' networks. However this parameter is not significant
for Croatia.

The negative and significant partner number of genres parameter
(for Romania only)
indicates that users who have friends who like many genres are less
likely to like jazz (while the positive and significant number of
genres parameter means that a user who likes many genres is more
likely to like jazz).

Estimating these models (200 parallel estimations) took approximately
21, 12, and 36 minutes, for Croatia, Hungary, and Romania,
respectively, on AMD EPYC 7543 2.8 GHz processors on a Linux
cluster. Less than 300 MB of memory per CPU was required for the Python code
running these estimations, and less than 8 GB in total was required
for the R scripts to do the final statistical computations and
diagnostic plots.

We also estimated the same models with ALAAMEE using the stochastic
approximation algorithm, and the results are shown in
\nameref{sitab:deezer_jazz_sa_estimations}. These estimations took
approximately 32, 14, and 36 hours (and less than 500 MB memory) for
Croatia, Hungary, and Romania, respectively, on the same system. The
results are consistent with those of the estimation with the EE
algorithm. However three parameter estimates were found to be
statistically significant by the stochastic approximation algorithm
that were not when using the EE algorithm: GWActivity for Croatia, and
GWActivity and TriangleT1 for Romania.

For the Romania network, the negative and significant TriangleT1
parameter indicates that users who like jazz are less likely to be
involved in a triangle structure than other users. This is evidence
that liking jazz is less likely to be observed in more clustered
regions of this network.

The TriangleT2 and TriangleT3 parameters need to be interpreted in
conjunction with Contagion and TriangleT1. The positive and
significant Contagion parameter indicates that liking jazz is socially
contagious, while the TriangleT1 parameter controls for the incidence
of liking jazz in closed triads. Given these effects in the model, a
positive and significant TriangleT2 parameter would indicate a type of
structural equivalence within a closed triad: a pair of users who are
friends and are both also friends with a common third user is
associated with that pair of users both liking jazz. A positive and
significant TriangleT3 parameter would indicate that jazz is socially
contagious within triads (over and above its contagion in dyads), as
reflected in an over-representation of closed triads where all three
users like jazz (given all the other effects in the model).  However
the TriangleT2 and TriangleT3 parameters are either not statistically
significant, or models containing them do not converge, in all three
countries, and so we cannot draw any conclusions about these
effects in this data.

Note that, as discussed in the Introduction, by using the ALAAM
model, in which the networks ties are fixed (exogenous to the model),
we are assuming that only the process of social contagion
is occurring, without the ability to account for the possibility that
friends (network neighbours) might both like the same genre of music
due to homophily.

Table~\ref{tab:deezer_alternative_ee_estimations} shows the results of
using the EE algorithm to estimate ALAAM parameters for the same three
networks, but this time with liking alternative music as the outcome
variable. For this data, the full model was able to be estimated for
all three countries. Degeneracy check plots for these models are
shown in \nameref{sifig:deezer_alternative_hr_degen_check},
\nameref{sifig:deezer_alternative_hu_degen_check}, and
\nameref{sifig:deezer_alternative_ro_degen_check}.

\begin{table}[!ht]
\begin{adjustwidth}{-2.25in}{0in} 
\centering
\caption{ {\bf ALAAM models for the Deezer networks, with liking ``alternative'' music as the outcome variable.}}
\begin{tabular}{|l|r|r|r|}
\hline
 Effect  & Croatia & Hungary & Romania\\ \thickhline  
Density  & $\heavy{\underset{(-3.364, -3.191)}{-3.277}}$ & $\heavy{\underset{(-3.111, -2.929)}{-3.020}}$ & $\heavy{\underset{(-3.262, -3.050)}{-3.156}}$\\ \hline
GWActivity [$\alpha = 2.0$]  & $\light{\underset{(-2.755, 2.064)}{-0.346}}$ & $\light{\underset{(-1.777, 1.954)}{0.088}}$ & $\light{\underset{(-2.039, 1.276)}{-0.381}}$\\ \hline
Contagion  & $\light{\underset{(-0.053, 0.032)}{-0.010}}$ & $\heavy{\underset{(0.095, 0.221)}{0.158}}$ & $\heavy{\underset{(0.028, 0.190)}{0.109}}$\\ \hline
TriangleT1  & $\light{\underset{(-0.007, 0.005)}{-0.001}}$ & $\light{\underset{(-0.022, 0.009)}{-0.006}}$ & $\light{\underset{(-0.054, 0.017)}{-0.018}}$\\ \hline
TriangleT2  & $\light{\underset{(-0.012, 0.017)}{0.003}}$ & $\light{\underset{(-0.032, 0.032)}{0.000}}$ & $\light{\underset{(-0.072, 0.062)}{-0.005}}$\\ \hline
TriangleT3  & $\light{\underset{(-0.046, 0.030)}{-0.008}}$ & $\light{\underset{(-0.057, 0.112)}{0.028}}$ & $\light{\underset{(-0.107, 0.195)}{0.044}}$\\ \hline
num. genres  & $\heavy{\underset{(0.438, 0.458)}{0.448}}$ & $\heavy{\underset{(0.408, 0.430)}{0.419}}$ & $\heavy{\underset{(0.393, 0.417)}{0.405}}$\\ \hline
partner num. genres  & $\light{\underset{(-0.002, 0.003)}{0.001}}$ & $\heavy{\underset{(-0.014, -0.006)}{-0.010}}$ & $\light{\underset{(-0.007, 0.003)}{-0.002}}$\\ \hline
\end{tabular}
\begin{flushleft} Parameter estimates are
    shown with 95\% confidence interval. Estimates that are
    statistically significant at the nominal $p < 0.05$ level are
    shown in bold. Results are from 200 converged runs (of a total 200).
\end{flushleft}
\label{tab:deezer_alternative_ee_estimations}
\end{adjustwidth}
\end{table}

The results are qualitatively quite similar to the model for jazz, in
that the only parameter estimate (other than Density) that is
statistically significant across all three networks is number of
genres, which is positive, and, again, the Contagion parameter is statistically
significant and positive for the Hungary and Romania (but not Croatia) networks. None
of the Triangle parameters are statistically significant in any of the
three countries.

Estimating these models (200 parallel estimations) took approximately
244, 74, and 47 minutes, for Croatia, Hungary, and Romania,
respectively (on the same system as the jazz model, and with the same
upper bounds on memory usage). We also attempted to estimate the same
models using the stochastic approximation algorithm, however for the
Croatia and Hungary networks these estimations did not complete within
a 48 hour elapsed time limit. The estimation for the Romania network
took approximately 30 hours (and less than 200 MB of memory), and the
resulting parameter estimates are shown in
\nameref{sitab:deezer_alternative_sa_estimations}. These estimates are
consistent with those estimated by the EE algorithm: the same three
parameters are found to be statistically significant with the same
sign, and all the other parameters are not statistically significant.

\section*{Conclusion}

ALAAMEE is open-source Python software for the estimation, simulation,
and goodness-of-fit testing of the ALAAM social contagion model. It
currently supports ALAAMs on undirected and directed one-mode
networks, and undirected two-mode (bipartite) networks. It also
supports estimation from snowball sampled network data.

Models can be estimated using either stochastic approximation with the
Robbins--Monro algorithm, or, for large networks, by the EE
algorithm. For networks small enough that it is practical
--- on the order of thousands of nodes, but also depending on the
model parameters and data --- we recommend using stochastic
approximation, a well-known and widely used method, which is
implemented in ALAAMEE just as it is in MPNet for ALAAM parameter
estimation. However for larger networks, stochastic approximation is
likely to be too slow to be practical, and so the EE algorithm can be
used instead. In this work we demonstrated its use on networks on
networks of approximately 50~000 nodes, for which stochastic
approximation was feasible in some, but not all, cases. The use of
the EE algorithm implemented in ALAAMEE to estimate ALAAM parameters
for a much larger network, with approximately 1.6 million nodes, for
which estimation by stochastic approximation is certainly not
feasible, is described in Stivala~\cite{stivala23b}.
This demonstrates that ALAAMEE allows estimation and goodness-of-fit
testing of ALAAM models for networks that are too large for prior
software implementations to estimate.

When using the EE algorithm, the results from the simulation experiments section indicate that it is always advantageous,
and does not result in a significantly increased false positive rate (at least up to the maximum of 500 tested in this work), to use as
many runs as practical, and 200 is probably more than sufficient in
most cases. Because these runs are entirely independent, elapsed time
can be minimized by running them in parallel using as many compute
cores and nodes as may be available.

The EE algorithm is fast, scalable to far larger networks, and able to
take advantage of multicore and large-scale parallel computing.
The simplified EE algorithm~\cite{borisenko19} used in ALAAMEE
has recently been shown, in the context of ERGMs, to be guaranteed
to converge to the MLE, if it exists, when the learning rate
(a parameter of the EE algorithm)
is sufficiently small~\cite{giacomarra23}. The proof uses the
uncertain energies framework of Ceperley \& Dewing~\cite{ceperley99}, as originally
suggested in Borisenko \etal~\cite{borisenko19}.

However, as shown in this work, using this algorithm with an insufficient number of estimation runs can result in
very low statistical power on some parameters, due to larger estimated
standard errors than the stochastic approximation method on the same
data. This has also been observed in using the EE algorithm to
estimate parameters for the ``citation ERGM'' (cERGM) ERGM variant~\cite{schmid21}, compared with using the default statnet MCMLE
algorithm~\cite{stivala22}.

Because ALAAMEE is written in Python using only the NumPy package, it
can easily be run on any system on which Python and NumPy are
available. This also facilitates its use in automated scripts, for
example for large-scale computational experiments such as those
described in this paper and in Stivala \etal~\cite{stivala20a},
which is not practical with a Windows application like MPNet.
A comparison of MPNet and ALAAMEE is shown in Table~\ref{tab:mpnet_comparison}.

\begin{table}[!ht]
\begin{adjustwidth}{-2.25in}{0in} 
\centering
\caption{
{\bf Comparison of MPNet and ALAAMEE software for ALAAM estimation.}}
\begin{tabular}{|l|l|l|l|}
\hline
\multicolumn{2}{|l|}{Feature} & MPNet & ALAAMEE \\ \thickhline
\multicolumn{2}{|l|}{Download location} & \url{http://www.melnet.org.au/pnet} & \url{https://github.com/stivalaa/ALAAMEE} \\ \hline
\multicolumn{2}{|l|}{Estimation algorithms} & Stochastic approximation (SA) & SA and equilibrium expectation (EE) \\ \hline
\multicolumn{2}{|l|}{Max.~network size (nodes)} & Thousands & Tens of thousands (SA, EE); $> 1.5$ million (EE only) \\ \hline 
\multicolumn{2}{|l|}{Platform}  & Microsoft Windows & Cross-platform (anywhere Python and R are available) \\ \hline
\multicolumn{2}{|l|}{Implementation language} & C\# & Python and R  \\ \hline
\multicolumn{2}{|l|}{User interface} & Windows GUI & Python functions \\ \hline
\multicolumn{2}{|l|}{Open source} & No & Yes \\ \hline
\multicolumn{2}{|l|}{User-extensible} & No & Yes \\ \hline
\multicolumn{2}{|l|}{Can use parallel computing} & No & Yes (EE algorithm only) \\ \hline
\multicolumn{2}{|l|}{First release year} & 2014 & 2020 \\ \hline
\multicolumn{2}{|l|}{Most recent release year} & 2022 & 2024 \\ \hline
\multicolumn{2}{|l|}{Snowball sampled networks} & No & Yes \\ \hline
\multirow{4}{5em}{Network types supported} & Undirected &  Yes & Yes \\ \cline{2-4}
& Directed & Yes &  Yes  \\ \cline{2-4}
& Two-mode & Yes  & Yes (currently undirected only) \\ \cline{2-4}
& Multilevel & Yes & No \\ \hline
\multirow{3}{5em}{Handles missing data} & Nodal attributes & No  & Simplistic (NA values ignored in statistic computation) \\ \cline{2-4}
& Outcome variable & No   & No (NA values mark outcome inapplicable in mode)    \\ \cline{2-4}
& Network ties & No  & No  \\ \hline
\end{tabular}
\begin{flushleft} Note: MPNet supports both ERGM and ALAAM, while ALAAMEE is only for ALAAM. MPNet features listed are those applicable to ALAAM.\\
\end{flushleft}
\label{tab:mpnet_comparison}
\end{adjustwidth}
\end{table}

The relative ease of Python programming, along with the simple and
internally documented implementations, with unit tests, of a variety
of ALAAM change statistics already included in the open-source ALAAMEE
software, should facilitate the creation of change statistics for any
new configurations required by users of the software.

A demonstration implementation of the EE algorithm for ERGM 
parameter estimation was written in
Python 
in the same style as ALAAMEE, that
is, using NumPy for linear algebra and Python dictionaries for graph
data structures. However this was found to be too slow for practical use,
and the software \cite{stivala_github_estimnetdirected}
was completely re-written in C~\cite{stivala20b}. For ALAAM, however,
where the MCMC process involves flipping binary variables in a vector
rather than edges in a graph, we found that the Python implementation
ALAAMEE was sufficiently fast for practical use not only with the EE
algorithm, but also with the Robbins--Monro algorithm.

In theory, ALAAMEE could be made considerably faster with little
effort by using a Python ``just-in-time'' (JIT) compiler, such as
Numba~\cite{lam15} or PyPy~\cite{bolz09}. However to date we have
been unable to accelerate ALAAMEE this way, finding that either the
code is not supported, or the ``accelerated'' version is actually
slower than the original.

In Parker \etal~\cite{parker22}, three limitations of ALAAMs are
identified. First, the outcome variable is restricted to binary;
second, the inherent assumption that the underlying social contagion
process is at equilibrium; and third, their inadequacy for very large
networks. We suggest that this work, in addition to the use of ALAAM
estimation from network snowball samples~\cite{stivala20a}
effectively addresses the third limitation. The second limitation can
be mitigated as suggested in Parker \etal~\cite{parker22}, by applying ALAAM
models to data in which the social network is observed at an
appropriate time before the outcome behavioural variable is. If
longitudinal (panel) data is available, in which the social network
ties and actor attributes of a population are observed at multiple
time points, the stochastic actor attribute model (SAOM) is a
more appropriate choice of model, enabling the estimation of
parameters corresponding to the co-evolution of the social network and
actor attributes~\cite{snijders17}.  This leaves the first
limitation, that the outcome variable must be binary, to be addressed
in future work. We have identified an additional limitation of ALAAMs
in this work and in Stivala~\cite{stivala23b}, namely that, like ERGMs, ALAAMs
can suffer from problems of ``near-degeneracy'' when only
simple statistics (such as Activity) are used, particularly
on larger networks. The use of the
GWActivity parameter can help overcome these problems, but as
illustrated by the examples in the ``\nameref{sec:results}'' section, it
still may not always be possible to fit a model with all the desired
parameters. Some avenues for further work on this problem are
suggested in Stivala~\cite{stivala23b}.

One further limitation of ALAAM estimation as implemented in ALAAMEE
is the handling of missing data. Missing nodal attributes can be
specified as ``NA'' values, in which case they are handled in a
simplistic way by simply not being counted towards the relevant change
statistics. For more than very small amounts of missing data, however,
it would be better to use some form of imputation. Missing values of
the outcome binary attribute may also be specified by ``NA'' values,
however in this case they act as ``structural zeroes'', which fix the
outcome variable at the NA value. This is appropriate in the case of
two-mode networks where the outcome variable is only defined for one
mode, and so it is set to NA for all nodes in the other mode. To
handle missing values of the outcome binary variable, the Bayesian
estimation method described by Koskinen \&
Daraganova~\cite{koskinen22} is more appropriate. Missing tie
variables are not handled at all, although it is possible to use
network snowball samples \cite{stivala20a}, which is also possible
with the Bayesian ALAAM estimation method \cite{koskinen22}.

An additional avenue of further work on ALAAM modeling was suggested in
the Introduction: the conception of an ALAAM as a bipartite
ERGM could be a way of implementing a multivariate ALAAM, that is, an
ALAAM with more than one outcome variable.

\section*{Supporting information}

\paragraph*{S1 Fig.}
\label{sifig:ipnet_meanse}
{\bf Parameter estimates and estimated standard errors from the
  stochastic approximation algorithm.}  The stochastic approximation
algorithm was used to estimate the known ALAAM parameters for the
Project 90 network with simulated attributes. This is the baseline
result for the Project 90 example in a study of the effect of network
sampling on ALAAM estimation \cite{stivala20a}. The error bars show
the nominal $95\%$ confidence interval. The horizontal line shows the
true value of the parameter, and each plot is annotated with the mean
bias, root mean square error (RMSE), the percentage of samples for
which the true value is inside the confidence interval (coverage
rate), and the Type II error rate (False Negative Rate, FNR). A
converged estimate was found for 99 of the total 100 samples.

\includegraphics[width=\textwidth]{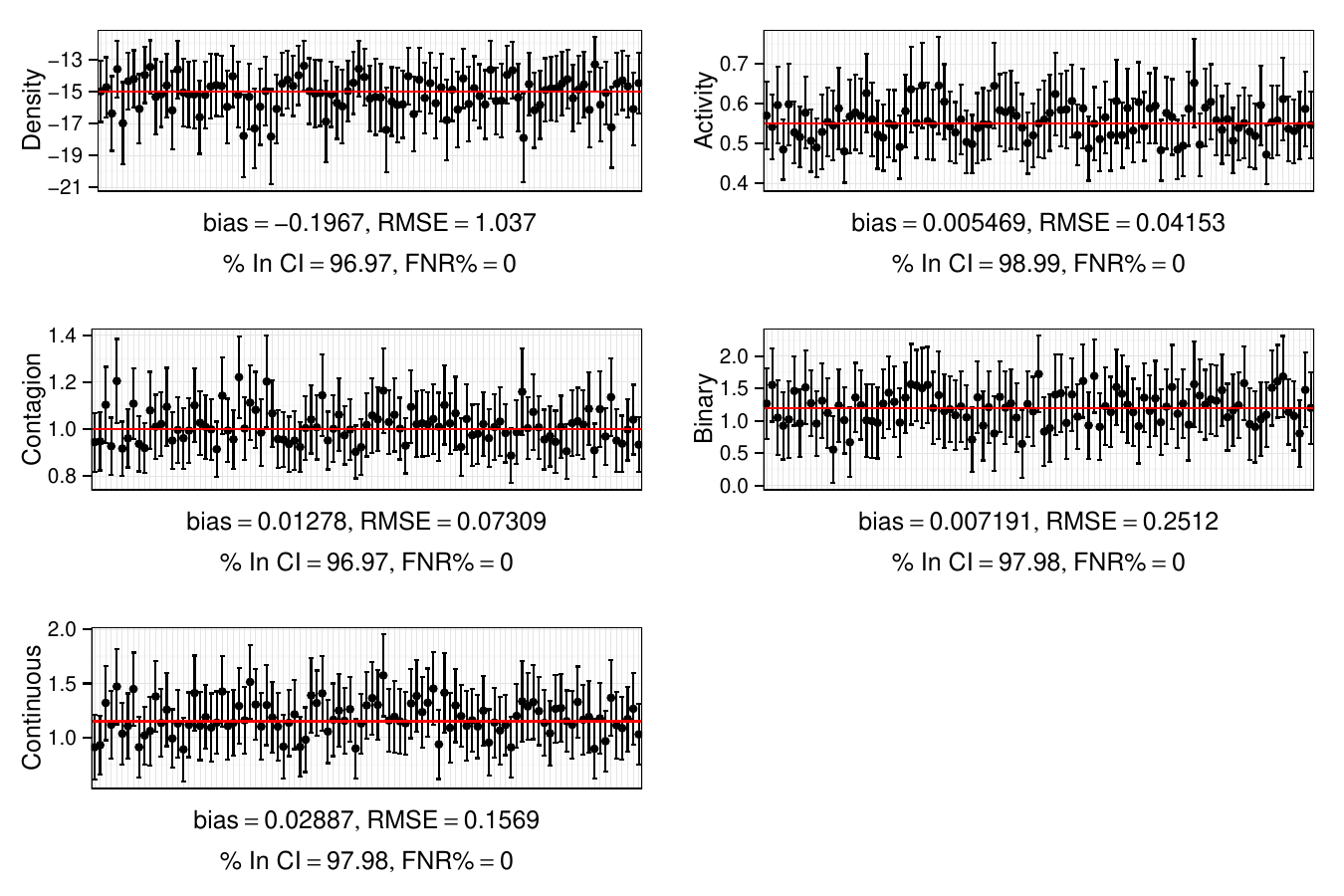}

\paragraph*{S2 Fig.}
\label{sifig:fpr_var_total_runs}
{\bf Type I error rate as the number of runs used for each sample is
varied.}

\includegraphics[width=\textwidth]{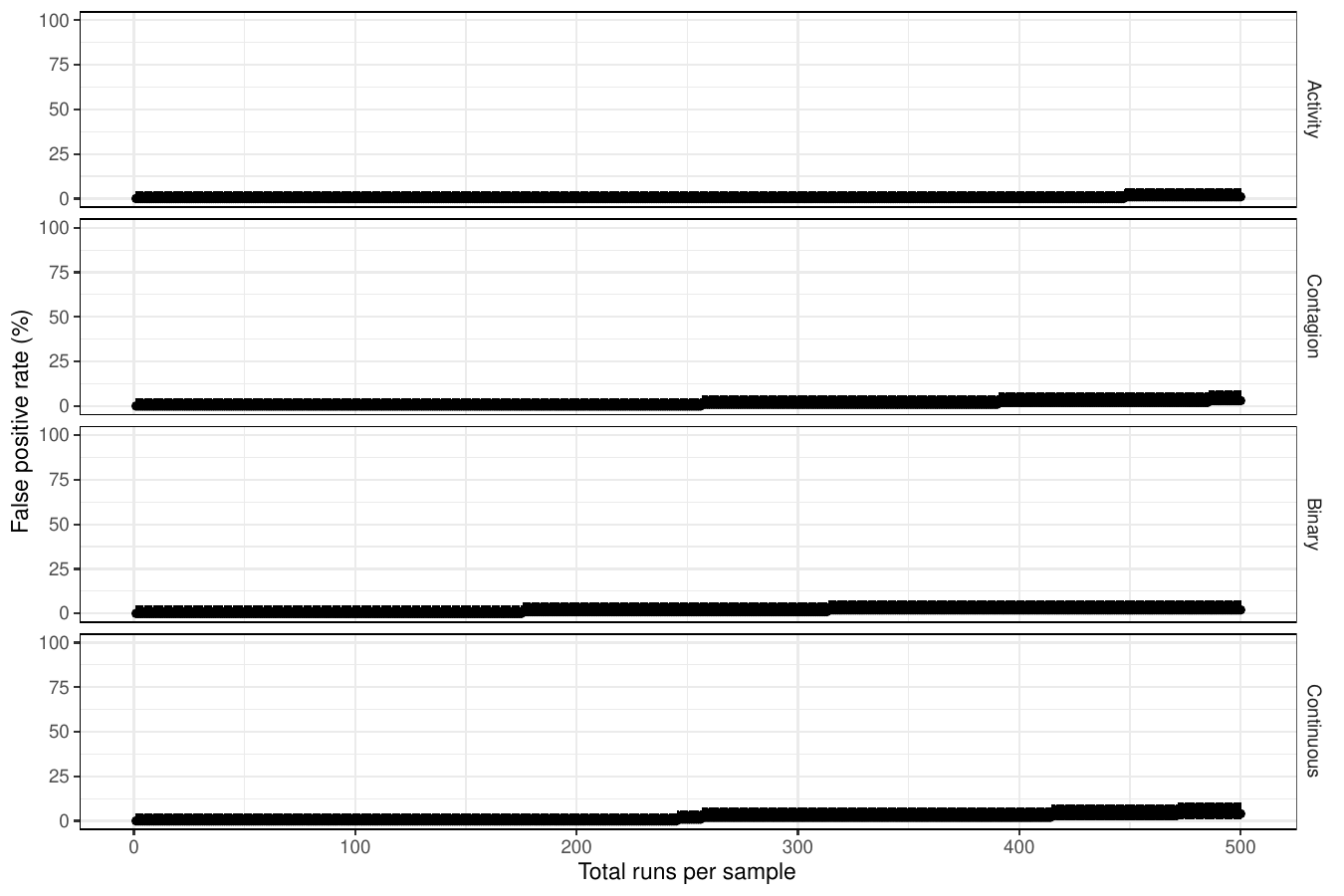}

\paragraph*{S3 Fig.}
\label{sifig:deezer_hr_degen_check} {\bf Degeneracy check for the
  Deezer Croatia network with liking jazz as the outcome variable.}
Trace plots and histograms show statistics of 100 networks simulated
from the model. The blue dashed lines on histograms show mean and blue
dotted lines on histograms and shaded areas on trace plots show 95\%
confidence interval, and red lines show the observed values.

\includegraphics[width=\textwidth]{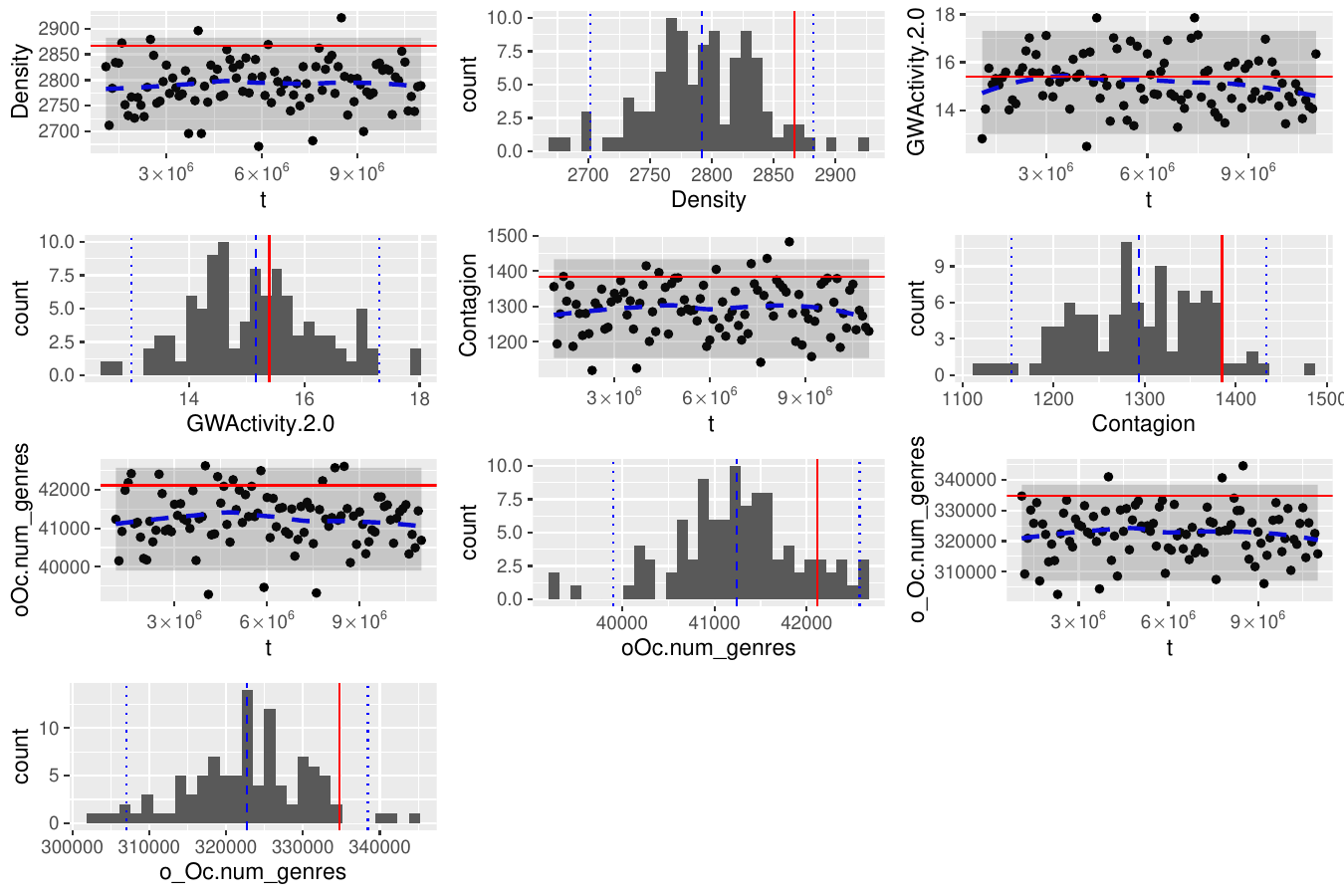}

\paragraph*{S4 Fig.}
\label{sifig:deezer_hu_degen_check} {\bf Degeneracy check for the
  Deezer Hungary network with liking jazz as the outcome variable.}
Trace plots and histograms show statistics of 100 networks simulated
from the model. The blue dashed lines on histograms show mean and blue
dotted lines on histograms and shaded areas on trace plots show 95\%
confidence interval, and red lines show the observed values.

\includegraphics[width=\textwidth]{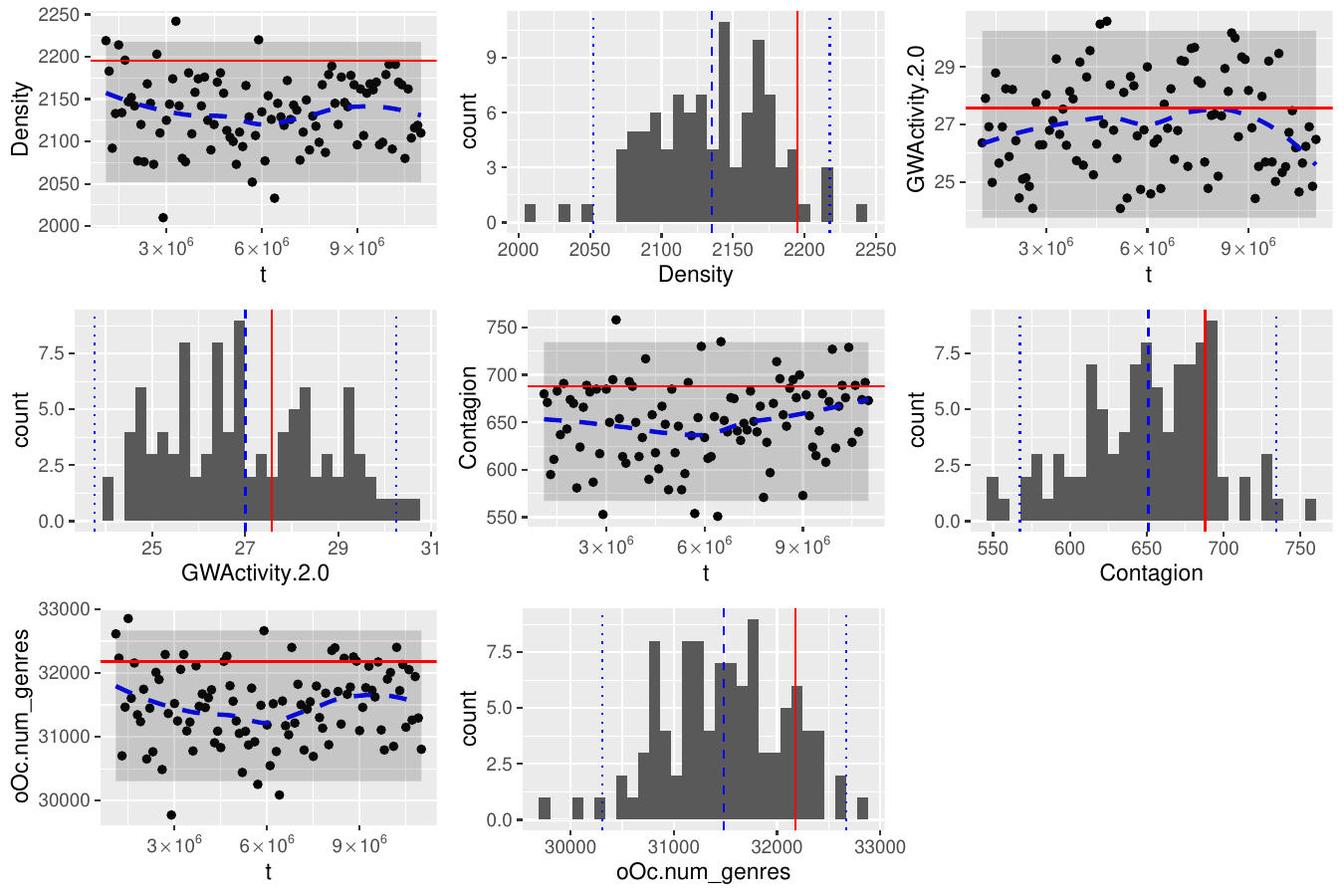}

\paragraph*{S5 Fig.}
\label{sifig:deezer_ro_degen_check} {\bf Degeneracy check for the
  Deezer Romania network with liking jazz as the outcome variable.}
Trace plots and histograms show statistics of 100 networks simulated
from the model. The blue dashed lines on histograms show mean and blue
dotted lines on histograms and shaded areas on trace plots show 95\%
confidence interval, and red lines show the observed values.

\includegraphics[width=\textwidth]{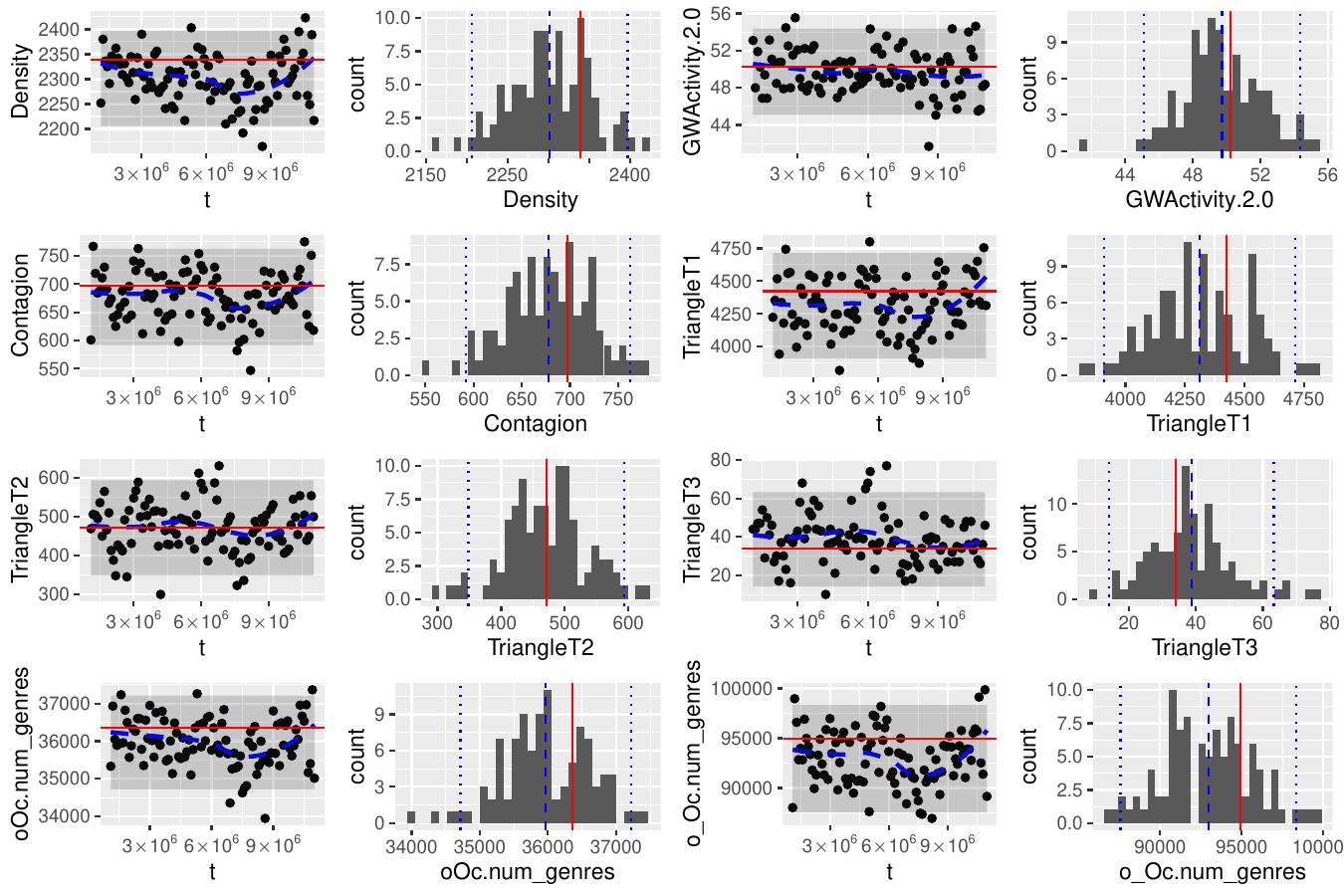}

\paragraph*{S1 Table.}
\label{sitab:deezer_jazz_sa_estimations}
{\bf Models estimated using with the stochastic approximation
algorithm, with liking jazz as the outcome
variable.} Parameter estimates are shown with their estimated standard
errors. Parameter estimates that are statistically significant at the
nominal $p < 0.05$ level are shown in bold.

  \begin{tabular}{lrrr}
    \hline
    Effect  & Croatia & Hungary & Romania\\
    \hline
    Density  & $\heavy{\underset{(0.050)}{-4.706}}$ & $\heavy{\underset{(0.051)}{-5.022}}$ & $\heavy{\underset{(0.057)}{-4.659}}$\\
    GWActivity [$\alpha = 2.0$]  & $\heavy{\underset{(0.861)}{-2.813}}$ & $\heavy{\underset{(0.681)}{5.266}}$ & $\heavy{\underset{(0.584)}{1.666}}$\\
    Contagion  & $\light{\underset{(0.032)}{-0.009}}$ & $\heavy{\underset{(0.030)}{0.266}}$ & $\heavy{\underset{(0.057)}{0.610}}$\\
    TriangleT1  & --- & --- & $\heavy{\underset{(0.009)}{-0.033}}$\\
    TriangleT2  & --- & --- & $\light{\underset{(0.049)}{0.009}}$\\
    TriangleT3  & --- & --- & $\light{\underset{(0.201)}{-0.068}}$\\
    num. genres  & $\heavy{\underset{(0.003)}{0.193}}$ & $\heavy{\underset{(0.003)}{0.192}}$ & $\heavy{\underset{(0.003)}{0.192}}$\\
    partner num. genres  & $\light{\underset{(0.000)}{-0.000}}$ & --- & $\heavy{\underset{(0.001)}{-0.006}}$\\
    \hline
  \end{tabular}

\paragraph*{S6 Fig.}
\label{sifig:deezer_alternative_hr_degen_check} {\bf Degeneracy check
  for the Deezer Croatia network with liking ``alternative'' music as
  the outcome variable.}  Trace plots and histograms show statistics
of 100 networks simulated from the model. The blue dashed lines on
histograms show mean and blue dotted lines on histograms and shaded
areas on trace plots show 95\% confidence interval, and red lines show
the observed values.

\includegraphics[width=\textwidth]{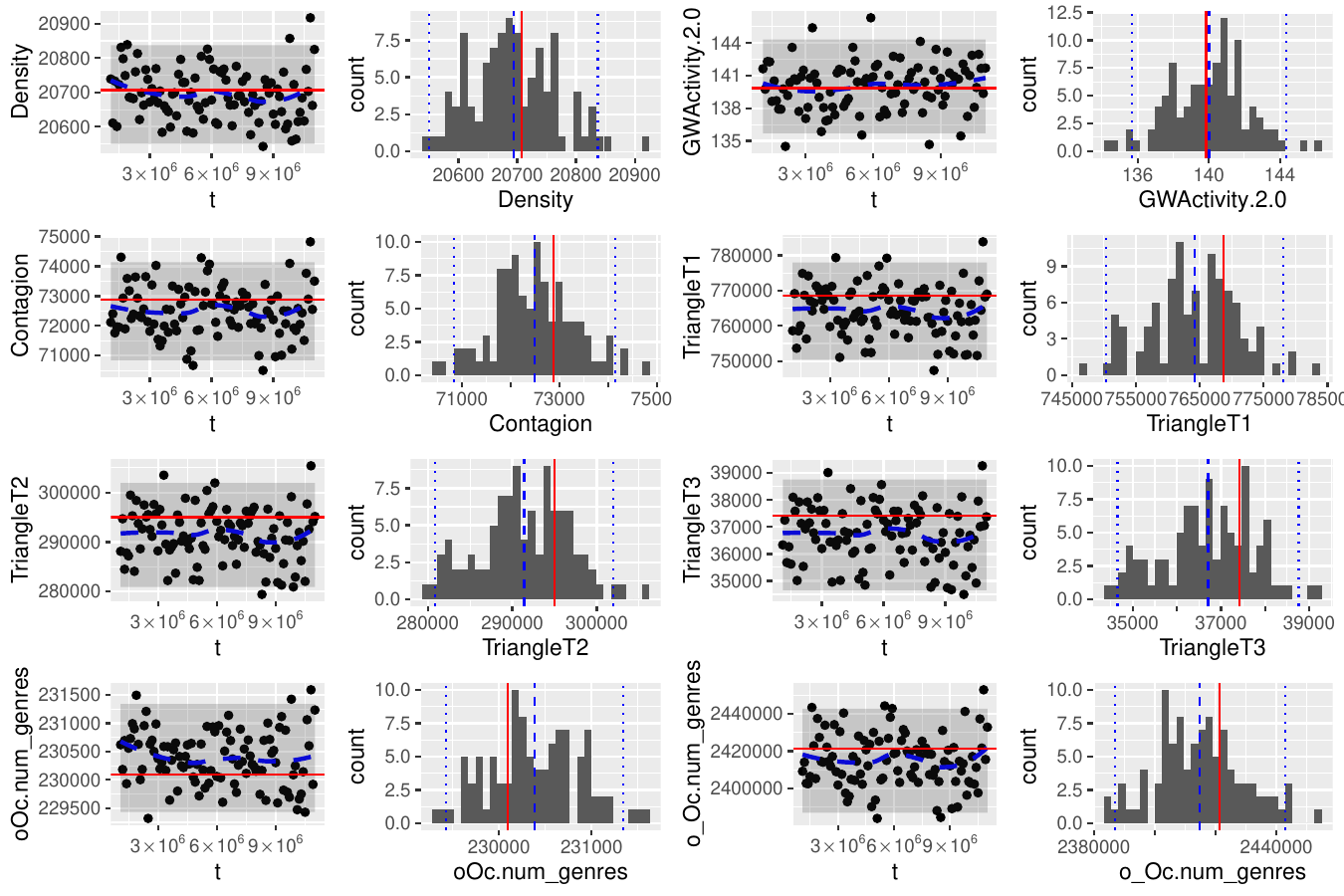}

\paragraph*{S7 Fig.}
\label{sifig:deezer_alternative_hu_degen_check} {\bf Degeneracy check
  for the Deezer Hungary network with liking ``alternative'' music as
  the outcome variable.}  Trace plots and histograms show statistics
of 100 networks simulated from the model. The blue dashed lines on
histograms show mean and blue dotted lines on histograms and shaded
areas on trace plots show 95\% confidence interval, and red lines show
the observed values.

\includegraphics[width=\textwidth]{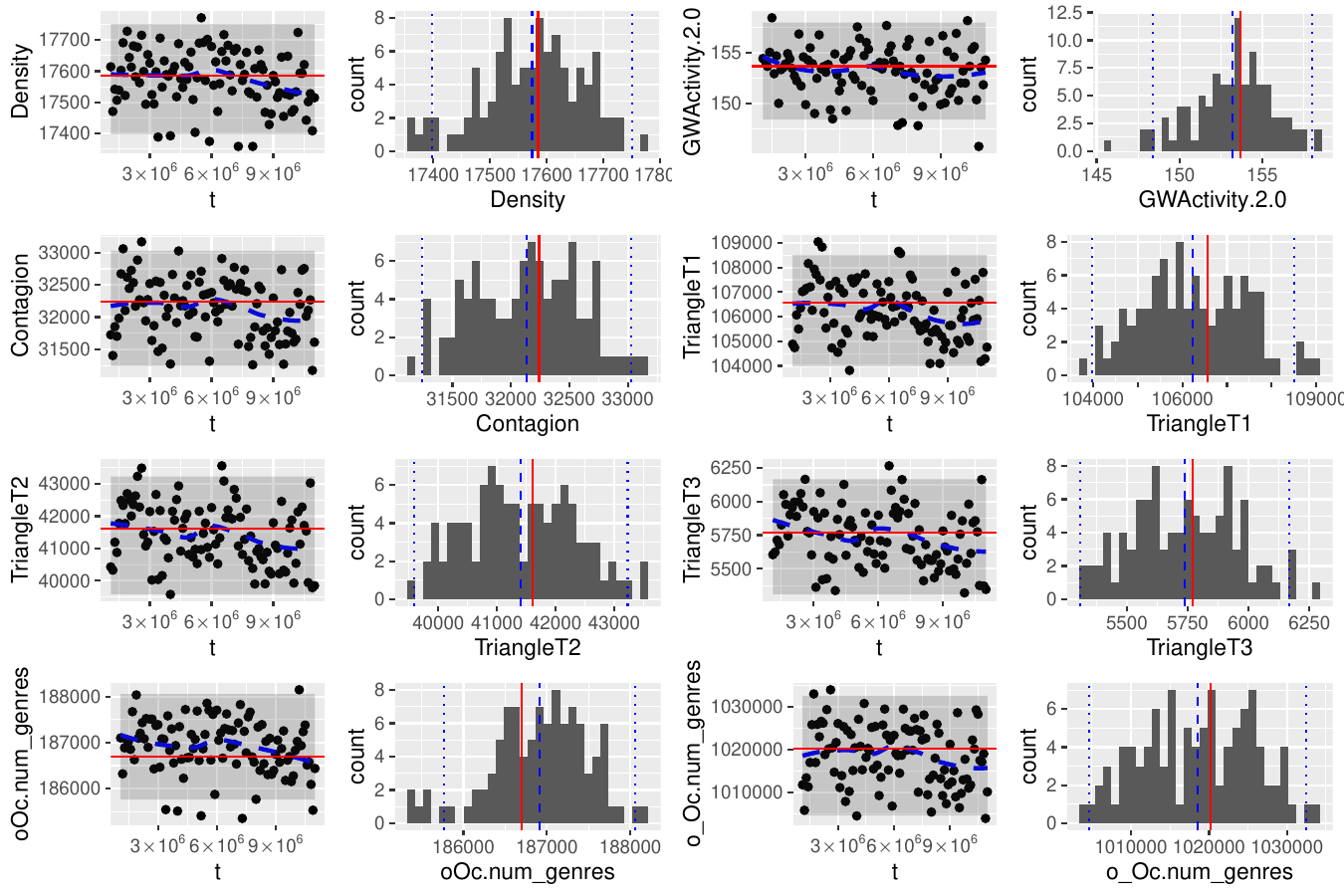}

\paragraph*{S8 Fig.}
\label{sifig:deezer_alternative_ro_degen_check} {\bf Degeneracy check
  for the Deezer Romania network with liking ``alternative'' music as
  the outcome variable.}  Trace plots and histograms show statistics
of 100 networks simulated from the model. The blue dashed lines on
histograms show mean and blue dotted lines on histograms and shaded
areas on trace plots show 95\% confidence interval, and red lines show
the observed values.

\includegraphics[width=\textwidth]{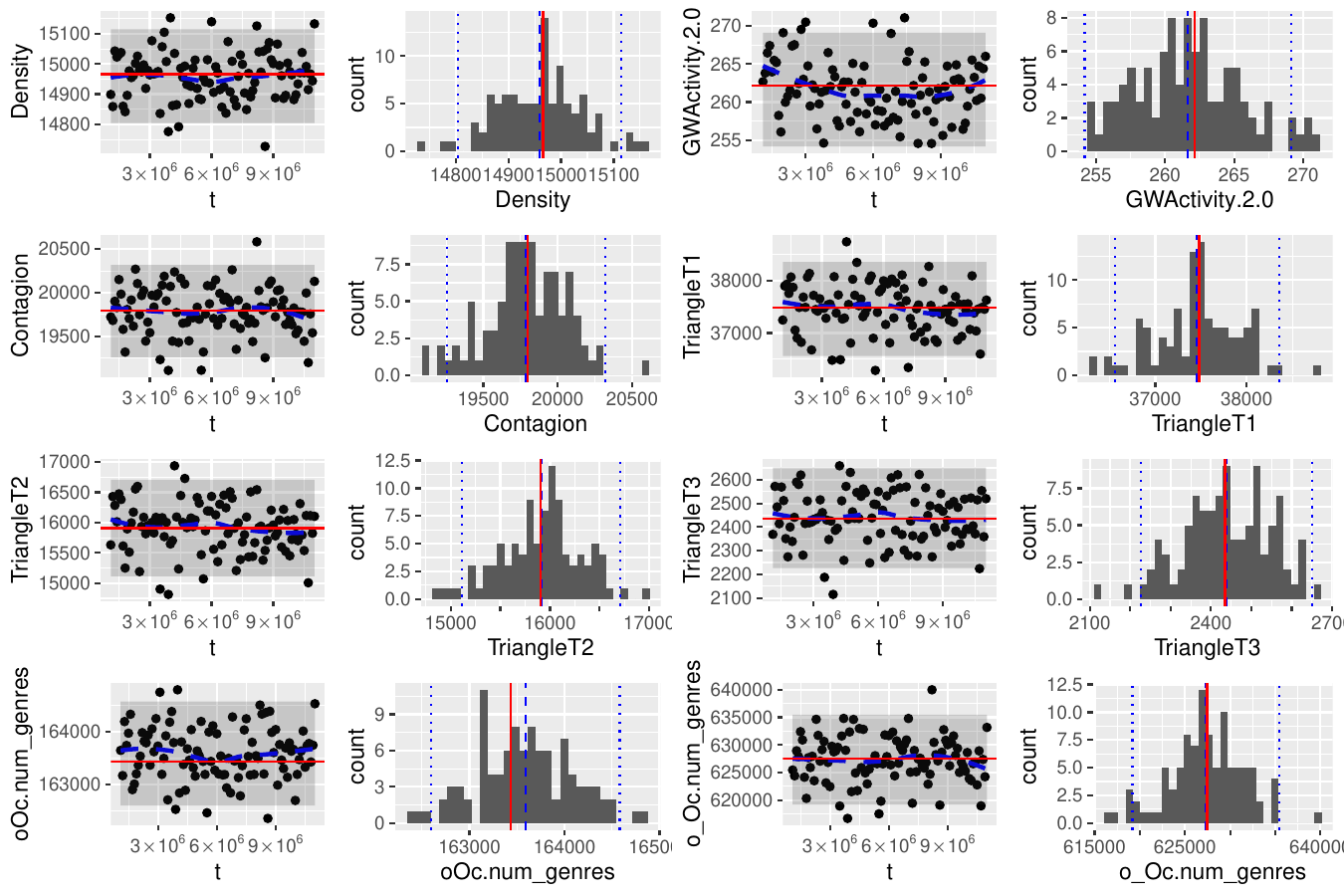}

\paragraph*{S2 Table.}
\label{sitab:deezer_alternative_sa_estimations}
{\bf Model estimated using with the stochastic approximation
algorithm, liking ``alternative'' music as the outcome
variable.}
Asterisks indicate statistical
significance at the nominal $p < 0.05$ level. Only the results for
Romania are shown, as estimation for the other two networks (Croatia
and Hungary) did not complete within the 48 hour elapsed time limit.

  \begin{tabular}{lrrrc}
    \hline
    Effect & Estimate & Std. error & t-ratio \\
    \hline
    Density & -3.136 & 0.032 & 0.015 & * \\ 
    GWActivity [$\alpha = 2.0$] & -0.352 & 0.334 & 0.016 &  \\ 
    Contagion & 0.110 & 0.019 & 0.031 & * \\ 
    TriangleT1 & -0.019 & 0.010 & -0.016 &  \\ 
    TriangleT2 & -0.004 & 0.021 & 0.018 &  \\ 
    TriangleT3 & 0.042 & 0.046 & 0.024 &  \\ 
    num. genres & 0.401 & 0.004 & -0.032 & * \\ 
    partner num. genres & -0.002 & 0.001 & 0.009 &  \\ 
    \hline
  \end{tabular}

\section*{Funding}

This work was funded by the Swiss National Science Foundation
(\url{https://www.snf.ch/en}) grant numbers 167326 (NRP 75) and 200778 to
AL. The funders had no role in study design, data collection and
analysis, decision to publish, or preparation of the manuscript.

\section*{Acknowledgments}

We used the high performance computing cluster at the Institute of
Computing, Universit\`a della Svizzera italiana
(\url{https://www.ci.inf.usi.ch/}), for initial computational experiments.
Computational experiments for the final manuscript were performed on the OzSTAR national facility at
Swinburne University of Technology
(\url{https://supercomputing.swin.edu.au/}). The OzSTAR program
receives funding in part from the Astronomy National Collaborative
Research Infrastructure Strategy (NCRIS,
\url{https://www.education.gov.au/ncris}) allocation provided by the
Australian Government, and from the Victorian Higher Education State
Investment Fund (VHESIF,
\url{https://www.vic.gov.au/projects-funded-victorian-higher-education-state-investment-fund})
provided by the Victorian Government.

\section*{Data availability statement}

The ``Project 90'' data
is
available upon registration from
\url{https://oprdata.princeton.edu/archive/p90/}. The excerpt of 50 girls
from the ``Teenage friends and lifestyle study'' data
is
available from
\url{https://www.stats.ox.ac.uk/~snijders/siena/s50_data.htm}. The
``Deezer'' data is available from the Stanford large
network dataset collection at
\url{https://snap.stanford.edu/data/gemsec-Deezer.html}.
All other data, source code, and scripts are freely
available from \url{https://github.com/stivalaa/ALAAMEE}.


%
%
%


\end{document}